\documentclass[journal]{IEEEtran}
\usepackage{subfigure}
\usepackage{booktabs}
\usepackage{algorithm}
\usepackage{algorithmic} 
\usepackage{graphicx}
\usepackage{picinpar}
\usepackage{mathrsfs}
\usepackage{amsmath,amssymb}
\usepackage{url}
\usepackage{flushend}
\usepackage[latin1]{inputenc}
\usepackage{colortbl}
\usepackage{soul}
\usepackage{multirow}
\usepackage{pifont}
\usepackage{color}
\usepackage{alltt}
\usepackage{enumerate}
\usepackage{siunitx}
\usepackage{breakurl}
\usepackage{epstopdf}
\usepackage{pbox}
\usepackage{multicol} 
\usepackage{bm}
\usepackage{changepage}    

\newtheorem{theoremx}{Theorem}
\newtheorem{lemmax}{Lemma}
\newtheorem{corollaryx}{Corollary}
\newtheorem{remark}{Remark}
\newtheorem{assumption}{Assumption}
\usepackage{graphicx}          

\begin{document}
\title{Abnormal source identification for parabolic distributed parameter systems}

	
	%
	%
	\author{
		\vskip 1em
		
		Yun Feng and Han-Xiong Li, \emph{Fellow, IEEE}
		
		\thanks{

			This work was supported in part by the General Research Fund Project
			from Research Grant Council of Hong Kong SAR (CityU: 11210719), in part by the National Natural Science Foundations of China under Grant U1501248, and a project from City University of Hong Kong (7005092). \emph{(Corresponding author: Han-Xiong Li.)}
			
            Y. Feng is with the Department of Systems Engineering and Engineering
            Management, City University of Hong Kong, Kowloon, Hong Kong,
            and also with the State Key Laboratory of High Performance Complex
            Manufacturing, Central South University, Changsha, 410083, P. R. China (e-mail:
            yun.feng@my.cityu.edu.hk) 
			
            H.-X. Li is with the Department of Systems Engineering and Engineering
            Management, City University of Hong Kong, Kowloon, Hong Kong (e-mail:
            mehxli@cityu.edu.hk).
			
		}
	}

	%



	\maketitle
	\begin{abstract}
    Identification of abnormal source hidden in distributed parameter systems (DPSs) belongs to the category of inverse source problems. It is important in industrial applications but seldom studied. In this paper, we make the first attempt to investigate the abnormal spatio-temporal (S-T) source identification for a class of DPSs. An \textbf{inverse S-T model} for abnormal source identification is developed for the first time. It consists of an adaptive state observer for source identification and an adaptive source estimation algorithm. One major advantage of the proposed inverse S-T model is that only the system output is utilized, without any state measurement. Theoretic analysis is conducted to guarantee the convergence of the estimation error. Finally, the performance of the proposed method is evaluated on a heat transfer rod with an abnormal S-T source.  
	\end{abstract}
	

	%
	\IEEEpeerreviewmaketitle

	\section{Introduction}
	Industrial processes such as thermal processes and transport-reaction processes can all be modeled by DPSs, whose system input, output, and parameters can change in both time and space domain, which can be concluded as the S-T dynamics~\cite{li2010modeling,wang2018incremental,wang2019reinforcement,wang2019dissimilarity,Xu2019,wang2018sliding,wang2019spatial,meng2018evolutionary}. Abnormal behaviors or events in DPSs may cause the failure of controller or undesired system response, both are harmful to the safe and reliable operation of the system. Without loss of generality, these abnormal behaviors or events can be considered as the result of an unknown abnormal S-T source $f(z,t)$ in the system dynamics, which can also be treated equivalently as S-T fault in the process. The abnormal source term $f(z,t)$ is a set of unknown terms which may cause undesirable system behaviors, including faults (actuator fault or sensor fault) occurring to the system, disturbance or noise coupling in the system dynamics, etc. Identification of the abnormal source term  has potential applications in chemical processes monitoring~\cite{el2006integrated,el2007actuator}, fault diagnosis of lithium-ion batteries~\cite{wei2019lyapunov}, control of vibrating single-link flexible manipulator system~\cite{zhao2019boundary}. On one hand, detection and identification of the abnormal source of DPSs are important for industrial applications but have not been fully investigated; On the other hand, the disturbance observer-based control (DOBC) of nonlinear parabolic PDE systems was studied in~\cite{wu2016disturbance,wu2016finite,wang2016low}, where the disturbance was governed by a known ordinary differential equation (ODE)~\cite{wu2016disturbance,wu2016finite} or partial differential equation (PDE)~\cite{wang2016low} exosystem with unknown initial conditions. 
	
	The abnormal S-T source is similar to the source term in the inverse source estimating problems for the wave equation, where the authors introduced the modulating functions-based method~\cite{asiri2017modulating} to address it. However, this approach requires full state measurement, as well as the computation of the measurement's derivative with respect to time, which are difficult to realize for industrial applications~\cite{Fischer2018}. 
	
	Over the past few decades, the actuator/sensor fault detection and diagnosis of DPSs have attracted more and more attention and some research efforts have been made, see~\cite{demetriou2002model,el2006integrated,el2007actuator,demetriou2007adaptive,armaou2008robust,ghantasala2009robust}. However, the fault detection and diagnosis of DPSs are inherently less complex than the abnormal S-T source detection and identification, due to the fact that the spatial distribution characteristic of the actuator/sensor fault was not considered: 
	\begin{itemize}
		\item In the actuator fault case, the spatial distribution function of the fault was assumed to be the same as that of the actuator and was known a prior;
		\item In the sensor fault case, the spatial distribution characteristic of the fault was not considered for the most commonly used point-wise measurement sensors.
	\end{itemize}
	Hence the actuator/sensor fault detection and diagnosis of DPSs are conducted only on the time domain, while the abnormal S-T source detection and identification are conducted on both the time and space domain. Existing approaches for fault detection and diagnosis of DPSs can be roughly classified into two categories: one was using a finite-dimensional ODE representation of DPSs; the other was based on the original PDE system. For example, the authors proposed a finite-dimensional residual generators for the purpose of fault detection of linear DPSs in~\cite{deutscher2016fault}. A novel model-based fault detection approach is developed in \cite{cai2016model}, where the state observer was based on the original PDE system. By using the modulating functions-based approach~\cite{shinbrot1954analysis} on the original PDE system, the actuator or sensor fault identification was derived by an algebraic expression in~\cite{fischer2016algebraic,fischer2017fault,fischer2018modulating}. Despite these innovative results, the studies on abnormal S-T source detection and identification of DPSs are relatively insufficient, which are of great significance from the application viewpoint. 
	
	Recently, the abnormal S-T source detection of DPSs was investigated using data-driven approaches for engineering applications~\cite{feng2018detection,feng2019dynamic}. However, the abnormal S-T source identification problem, i.e. estimating the unknown source term $f(z,t)$, was not considered, which is significant in abnormal source removing and process restoring. Moreover, compared to abnormal S-T source detection, the abnormal S-T source identification is more involved since it requires the dynamic tracking of the abnormal S-T source rather than determining detection thresholds for the generated residuals~\cite{feng2018detection,feng2019dynamic}.

	Adaptive observers~\cite{wang1996actuator,jiang2002adaptive,jiang2006fault,zhang2008adaptive} are efficient tools for the identification of unknown terms in dynamical systems modeled using ODEs. However, owing to the infinite-dimensional characteristic of DPSs, the adaptive observers' design methodologies for ODE systems cannot be applied to DPSs directly whose dynamical behaviors are modeled using PDEs. As one of the most representative classes of DPSs, a parabolic DPS's spatial differential operator can be divided into a finite-dimensional slow subsystem and an infinite-dimensional fast subsystem~\cite{christofides2012nonlinear}. This characteristic of parabolic DPSs provides a potential path of applying the adaptive observers for the abnormal S-T source identification problem.  
	
	The advantage of using adaptive observers for the abnormal S-T source identification problem over modulating-functions approaches~\cite{asiri2017modulating} include:
	
	\begin{itemize}
		\item First, no state measurement is needed, so fewer sensors are required in industrial applications; 
		\item Second, neither the time or space derivative of the measurement is required, which will improve the identification performance.  
	\end{itemize} 
	
	Motivated by the above-mentioned considerations, we make the first attempt to investigate the abnormal S-T source identification problem for a class of linear parabolic DPSs in this paper. An inverse S-T model for abnormal source identification is developed based on the theorem of separation of variables. The inverse S-T model consists of an adaptive state observer for source identification and an adaptive source estimation algorithm. Both theoretical analysis and numerical simulations are provided to validate the feasibility and effectiveness of the proposed method.             
	
	The rest of this paper can be summarized as follows: The problem descriptions are provided in Section~\ref{sec:Preliminaries and problem statement}. In Section~\ref{sec:Adaptive fault diagnosis observer design}, the inverse S-T model design for abnormal S-T source identification is presented. Theoretic analysis and numerical simulations are provided in Section~\ref{sec:teeoretic analysis} and Section~\ref{sec:Numerical Simulation}, respectively. Finally, this paper is concluded in Section~\ref{sec:Conclusion}.

	\section{Preliminaries and problem statement}\label{sec:Preliminaries and problem statement}
	\subsection{System description}
	\begin{figure}[!h]
		\centering
		\includegraphics[width=5cm]{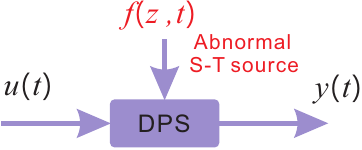}
		\caption{System description.}
		\label{fig:problem}
	\end{figure}
	Consider a class of linear DPSs modeled by the following parabolic PDE:
	\begin{equation} \label{e1}
	\begin{aligned}
	\frac{{\partial x(z,t)}}
	{{\partial t}} &= {a_1}\frac{{\partial x(z,t)}}
	{{\partial z}} + {a_2}\frac{{{\partial ^2} {x(z,t)}}}
	{{\partial {z^2}}}+a_3x(z,t)\\ 
	&+ {k_u}\bm{b}_u^T(z)\bm{u}(t) + f(z,t), \hfill \\
	\bm{y}(t) &= \int_{{\alpha _1}}^{{\alpha _2}} {\bm{c}(z)k_y x(z,t)dz},  \hfill \\ 
	\end{aligned} 
	\end{equation}
	
	subject to the following boundary conditions:
	\begin{equation} \label{e2}	
	\begin{gathered}
	c_1 {x}({\alpha _1},t)+d_1\frac{{\partial {x}}}
	{{\partial z}}({\alpha _1},t) = {r_1}, \hfill \\
	c_2 {x}({\alpha _2},t)+d_2\frac{{\partial {x}}}
	{{\partial z}}({\alpha _2},t) = {r_2}, \hfill \\ 
	\end{gathered} 
	\end{equation}
	
	and with the following initial condition:
	\begin{equation} \label{e3}
	x(z,0) = {{{x}}_0}(z),
	\end{equation}
	
	where $x(z,t)$ denotes the state variable; $\left[ {{\alpha _1},{\alpha _2}} \right] \subset \mathbb{R}$ is the spatial domain of the system; $z \in \left[ {{\alpha _1},{\alpha _2}} \right]$ is the spatial coordinate; $t \in \left[ {0,\infty } \right)$ denotes the time; $\bm{u}(t) \in {\mathbb{R}^{{n_u}}}$ denotes the vector of manipulated input; $f(z,t) \in \mathbb{R}$ denotes the unknown abnormal S-T source in the DPS, which is the root of abnormal behaviors or events. The definition \textbf{``abnormal source''} is proposed to distinguish from the manipulated input $\bm{u}(t)$, which can be considered as a \textbf{``normal source''}. The abnormal S-T source  $f(z,t)$ is irrelevant to the state variable $x(z,t)$ and can't be measured; $\bm{y}(t) \in {\mathbb{R}^{{n_y}}}$ denotes the system output; $\partial x/\partial z$ and ${\partial ^2}x/\partial {z^2}$ denote the first and second-order spatial derivatives of $x$, respectively; $a_1, a_2, a_3, k_u, k_y, c_1, c_2, d_1, d_2, r_1$, and $r_2$ are constant coefficients; The $i$th element of known smooth function $\bm{b}_u(z)\in \mathbb{R}^{n_u}$ describes how the $i$th element of control action $\bm{u}(t)$ is distributed in $\left[ {{\alpha _1},{\alpha _2}} \right]$; The $i$th element of known smooth function $\bm{c}(z)\in \mathbb{R} ^ {n_y}$ is determined by the shape (point or distributed) of the $i$th measurement sensor; And ${{{x}}_0}(z)$ is the initial condition. A schematic graph of the system description is shown in Fig.~\ref{fig:problem}.

	The following Hilbert Space is defined throughout this paper:
	\begin{equation*}
	{\mathcal{H}}\mathop  = \limits^\Delta \mathcal{L}_2([{\alpha _1},{\alpha _2}];\mathbb{R}). 
	\end{equation*} 
	Also, in this Hilbert Space, we have the following inner product and norm: 
	\begin{equation*}
	\begin{aligned}
	< {{ {x}}_1(\cdot)},{{{x}}_2(\cdot)} > \mathop  = \limits^\Delta  \int_{{\alpha _1}}^{{\alpha _2}} {{ {x}}_1(z)} { {x}}_2(z)dz,\\
	{\left\| { {x_1(\cdot)}} \right\|_{2}}\mathop  = \limits^\Delta < {{ {x}}_1(\cdot)},{{{x}}_1(\cdot)} >  ^{1/2},\\
	\end{aligned}
	\end{equation*}    
	where ${x}_1(\cdot)$ and ${x}_2(\cdot)$ are two elements of ${\mathcal{H}}$.

	\subsection{Problem statement}
	The problem investigated in this paper can be summarized as follows: 
	
	\begin{adjustwidth}{2em}{2em}
		Utilize the system output $\bm{y}(t)$ to design an inverse S-T model to identify the abnormal S-T source ${f}(z,t)$ of systems, 
	\end{adjustwidth}

	which subject to the system model in (\ref{e1}) with the boundary conditions in (\ref{e2}) and the initial condition in (\ref{e3}).   
	
	\section{Inverse S-T model design for abnormal S-T source identification}\label{sec:Adaptive fault diagnosis observer design}
	In this section, an inverse S-T model is developed to infer the unknown source term $f(z,t)$ from the system output $\bm{y}(t)$, without any state measurement. 
	
	\begin{figure}[!h]
		\centering
		\includegraphics[width=5cm]{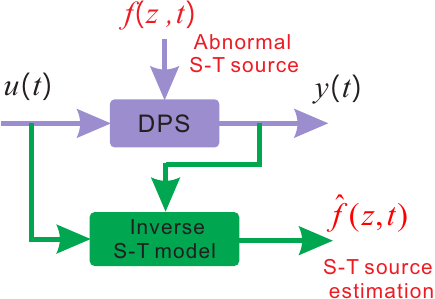}
		\caption{Framework of the inverse S-T model.}
		\label{fig:methodology}
	\end{figure}
	
	Motivated by the adaptive observer theory~\cite{wang1996actuator,jiang2002adaptive,jiang2006fault,zhang2008adaptive}, the inverse S-T model in Fig.~\ref{fig:methodology} is developed based on an adaptive state observer for source identification. 
	
	Considering the infinite-dimensional characteristic of the linear DPSs described in (\ref{e1})-(\ref{e3}), an approximate finite-dimensional system model which exhibits the dominant dynamics of the original system is first derived for the adaptive state observer design as follows:
	\begin{equation} \label{e11}
	\begin{aligned}
	{{\dot {\bm{x}}}_s}(t) &= {{\bm{A}}_s}{\bm{x}_s}(t)+ {\bm{B}_{u,s}}\bm{u}(t)+\bm{f}_s(t), \hfill \\
	{\bm{y}}_s(t) &= {\bm{C}_{s}{\bm{x}_s}(t)}, \hfill \\
	\end{aligned}
	\end{equation} 
	The advantages include:
	\begin{itemize}
		\item Finite-dimensional observers can then be applied to the abnormal S-T source identification rather than infinite-dimensional ones, by avoiding extra design efforts.
	\end{itemize} 
	In industrial applications, dominant models are sufficient for satisfactory performance without extra design efforts. Details of model reduction can be referred to the Appendix. 
	
	\begin{remark}\textbf{Time/space decoupled form of the abnormal S-T source}\\
		\emph{
		From the definition of $\bm{f}_s(t)$ and $\bm{f}_f(t)$ in (\ref{e8}), it is easy to obtain that: 
		\begin{equation*}
		f(z,t) = {\bm{\phi} ^T}(z)\bm{f}(t) = \bm{\phi} _s^T(z){\bm{f}_s}(t) + \bm{\phi} _f^T(z){\bm{f}_f}(t).
		\end{equation*}
		In this manner, the unknown abnormal S-T source can be described by the products of the basis functions ${\bm{\phi}}(z)$ and the temporal coefficients $\bm{f}(t)$. Hence the abnormal S-T source identification is transformed into the identification of the temporal coefficients $\bm{f}(t)$.}    
	\end{remark}

	Motivated by the adaptive fault diagnosis observer introduced in~\cite{wang1996actuator,jiang2002adaptive,jiang2006fault,zhang2008adaptive}, an adaptive state observer for (\ref{e11}) is constructed as follows:
	\begin{equation} \label{e12}
	\begin{aligned}
	{{\dot {\hat{\bm{x}}}_s}(t)} &= {{\bm{A}}_s}{\hat{\bm{x}}_s}(t)+ {\bm{B}_{u,s}}\bm{u}(t)+\hat{\bm{f}}_s(t)-\bm{L}(\hat{\bm{y}}_s(t)-\bm{y}(t)), \hfill \\
	{\hat{\bm{y}}}_s(t) &= {\bm{C}_{s}{\hat{\bm{x}}_s}(t)}, \hfill \\
	\end{aligned}
	\end{equation} 
	where $\hat{{\bm{x}}}_s(t) \in \mathbb{R}^m$, $\hat{\bm{y}}_s(t)\in \mathbb{R}^{n_y}$, and $\hat{\bm{f}}_s(t)\in \mathbb{R}^{m}$ are the estimates of ${{\bm{x}}}_s(t)$, ${{\bm{y}}}_s(t)$, and ${{\bm{f}}}_s(t)$, respectively. $\bm{L}\in \mathbb{R}^{m \times n_y}$ is the observer gain matrix. 
	
	Define $\bm{e}_x(t)={\hat{\bm{x}}_s}(t)-{\bm{x}_s}(t)$, $\bm{e}_y(t)={\hat{\bm{y}}_s}(t)-{\bm{y}}(t)$, and $\bm{e}_f(t)={\hat{\bm{f}}_s}(t)-{\bm{f}_s}(t)$, then the error dynamics are obtained by combing (\ref{e11}) and (\ref{e12}):
	\begin{equation} \label{e13}
	\begin{aligned}
	{{\dot {{\bm{e}}}_x}(t)} &= ({{\bm{A}}_s}-\bm{L}\bm{C}_s){{\bm{e}}_x}(t)+{\bm{e}}_f(t)+\bm{L}\bm{y}_f(t), \hfill \\
	{{\bm{e}}}_y(t) &= {\bm{C}_{s}{{\bm{e}}_x}(t)}-\bm{y}_f(t). \hfill \\
	\end{aligned}
	\end{equation} 
	
	\begin{remark}\emph{
		To be noticed, the output error $\bm{e}_y(t)$ which can be used for the adaptive state observer design is defined as:
		\begin{displaymath}
		\bm{e}_y(t)={\hat{\bm{y}}_s}(t)-{\bm{y}}(t)
		\end{displaymath}
		rather than the error between the slow-system output estimation ${\hat{\bm{y}}_s}(t)$ and the slow system output ${\bm{y}}_s(t)$. The reason is that the slow-system output ${\bm{y}}_s(t)$ in (\ref{e11}) can not be directly obtained while the original PDE system output $\bm{y}(t)$ in (\ref{e1}) can be obtained by the measurement sensors. The problem that comes with it is that $\bm{e}_y(t)\ne{\bm{C}_{s}{{\bm{e}}_x}(t)}$, which is equal in ~\cite{wang1996actuator,jiang2002adaptive,jiang2006fault,zhang2008adaptive}. Therefore, as shown in (\ref{e13}), the output of fast subsystem $\bm{y}_f(t)$ should be considered in the adaptive state observer design, which will introduce extra errors for the source identification. One make-up solution in practice is to select a sufficient high order $m$ for the slow subsystem.}     
	\end{remark}
	Then the following adaptive source estimation algorithm is proposed:
	\begin{equation}\label{e17}
	{\dot {\hat{\bm{f}}}}_s(t) =  - \bm{\Gamma F}({\dot{\bm{e}}_y}(t)+\sigma{{\bm{e}}_y}(t)),
	\end{equation}
	where $\bm{F}\in \mathbb{R} ^ {m \times n_y}$ and the symmetric positive definite matrix $\bm{\Gamma}\in \mathbb{R} ^ {m \times m}$ is the learning rate. 
	\begin{remark}\emph{
		It can be found that the adaptive source estimation algorithm (\ref{e17}) consists of a proportional term and an integral one of ${\bm{e}}_y(t)$, it can be rewritten as:
		\begin{displaymath}
		{{\hat{\bm{f}}}}_s(t) =  - \bm{\Gamma F}({{\bm{e}}_y}(t)+\sigma\int_0^t {{\bm{e}_y}(\tau )} d\tau).  
		\end{displaymath}
		The proportional term is introduced to improve the rapidity of abnormal source estimation.}
	\end{remark}

	\begin{figure}[!h]
		\centering
		\includegraphics[width=9cm]{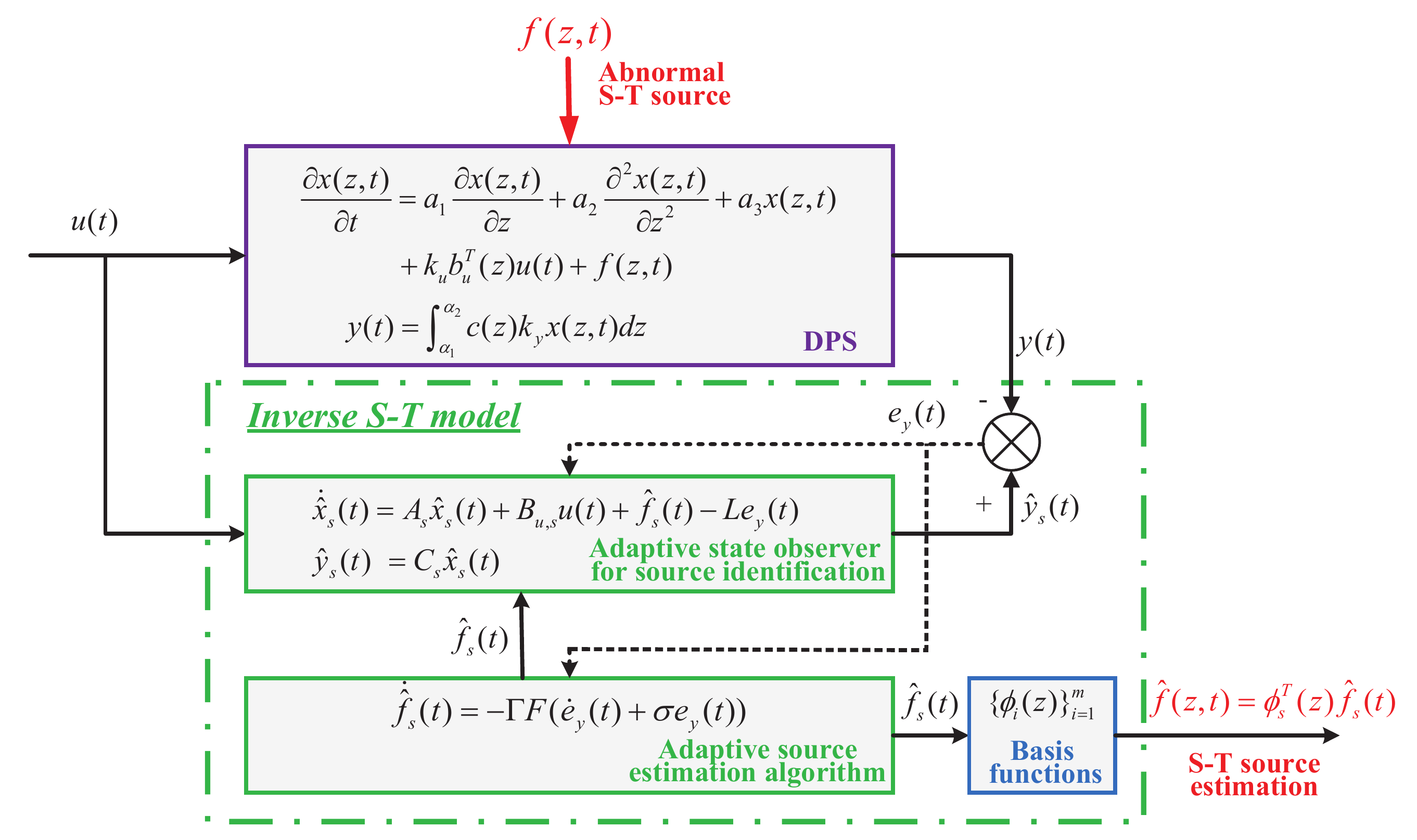}
		\caption{Schematic graph of the proposed inverse S-T model-based method for source identification.}
		\label{fig:schematic}
	\end{figure}
	
	Finally, the abnormal S-T source estimation can be done by the \textbf{time/space synthesis}:
	\begin{equation}
	\label{eqn:t/ssynthesis}
	\hat{f}(z,t)=\bm{\phi} _s^T(z){\hat{\bm{f}}_s}(t).  
	\end{equation}

	\begin{remark}\emph{
		To be noticed, a general assumption of the abnormal S-T source $f(z,t)$ is that it can be formulated in the following basis expansion form:
		\begin{displaymath}
		f(z,t)=\sum\limits_{i = 1}^n {{f_i}(t){\upsilon _i}(z)},\; 1\leqslant n<\infty 
		\end{displaymath}   
		with an finite number of the basis functions $\left\{ {{\upsilon _i}(z)} \right\}_{i = 1}^n$, where $n$ is unknown for the abnormal S-T source identification. Unlike the polynomial-type basis functions used in~\cite{asiri2017modulating}, the eigenfunctions $\left\{ {{\phi _i}(z)} \right\}_{i = 1}^n$ are used as the basis functions instead, i.e. ${\upsilon _i}(z) = {\phi _i}(z),i = 1,\cdots, n.$ When the order of the slow subsystem $m\ge n$, one can obtain that $\bm{f}_f(t)=0$. Since the inverse S-T model is based on an approximate finite-dimension model, $\bm{f}_f(t)$ is neglected in the abnormal S-T source identification. However, to further enhance the abnormal source identification performance, one can select a sufficient large $m$ under this assumption, which can also reduce the influence caused by the neglecting of the fast subsystem.}        
	\end{remark}

	A schematic graph of the proposed inverse S-T model-based method for abnormal S-T source identification is summarized in Fig.~\ref{fig:schematic}, it can be found that the inverse S-T model consists of the adaptive state observer for source identification in (\ref{e12}) and the adaptive source estimation algorithm in (\ref{e17}). To be noticed, in the proposed inverse S-T model, no state measurement is used.

	\section{Theoretical analysis}\label{sec:teeoretic analysis}
	The following assumptions and lemma are needed for the proposed inverse S-T model design.
	%

	  \begin{assumption}\emph{
			The output of the $\bm{x}_f$-subsystem $\bm{y}_f(t)$ and its derivative with respect to time ${{\dot{\bm{y}}_f}(t)}$ satisfy that:
			\begin{equation*}
			\begin{aligned}
			\left\| {{\bm{y}_f}(t)} \right\|_{peak} &\triangleq \sup _{t}| | {\bm{y}_f}(t)| |<\infty, \hfill \\
			{\left\| {{\dot{\bm{y}}_f}(t)} \right\|_{peak}} &\triangleq \sup _{t}| | {\dot{\bm{y}}_f}(t)| |<\infty,\; \forall t \geqslant 0 \hfill \\ 
			\end{aligned}
			\end{equation*} 
			where $\left\|  \cdot  \right\|_{peak}$ and $\left\|  \cdot  \right\|$ denote the  so-called peak-norm~\cite{ding2008model} and Euclidean norm, respectively.} 
	\end{assumption}

	
	\begin{assumption}\emph{
		The derivative of $\bm{f}_s(t)$ with respect to time is norm bounded, i.e. 
		\begin{equation*}
		\left\| {{\dot{\bm{f}}_s}(t)} \right\|^2 \leqslant {f_1},\;\forall t \ge 0
		\end{equation*} 
		where $f_1 \in \left[ {0,\infty } \right)$ is a constant.} 
	\end{assumption}
	
	\begin{assumption}\emph{
		The abnormal S-T source $f(z,t)$ satisfies that: 
		\begin{equation*}
		\left\| {f(z,t)} \right\|_2^2 = {\left\| {{\bm{f}_s}(t)} \right\|^2} + {\left\| {{\bm{f}_f}(t)} \right\|^2} \leqslant {f_2},\;\forall t \ge 0
		\end{equation*}
		where $f_2 \in \left[ {0,\infty } \right)$ is a constant.} 
	\end{assumption}
	
	\begin{assumption}\emph{
		$(\bm{A}_s,\bm{C}_s)$ is observable and $\bm{C}_s$ is of full column rank.}  
	\end{assumption}
	
	\begin{remark}\emph{
		In Assumption 4, the requirement of $\bm{C}_s$ being full column rank is common set in fault isolation, it is also known as the output separability condition~\cite{white1987detection,liu1997fault}. This can be done by selecting appropriate measurement sensors and dominant modes.}   
	\end{remark}
	
	\begin{lemmax} \cite{jiang2002adaptive}\emph{
		For a given positive scalar $\mu>0$ and a symmetric positive definite matrix $\bm{P}$, the following inequality holds:
		\begin{displaymath}
		2{\bm{x}^T}\bm{y} \leqslant \frac{1}{\mu }{\bm{x}^T}\bm{Px} + \mu {\bm{y}^T}{\bm{P}^{ - 1}}\bm{y}, \;\bm{x}, \bm{y} \in {\mathbb{R}^n}.
		\end{displaymath}}
	\end{lemmax}

    	\begin{lemmax} \cite{ioannou2012robust}\emph{
         Let $V(t)$ and $g(t)$ be real functions. Then
    		\begin{displaymath}
            \dot{V}(t) \leq-\alpha V(t)+g(t), \forall t \geq 0
    		\end{displaymath}
    		implies that
    		\begin{displaymath}
    		V(t) \leq e^{-\alpha t} V(0)+\int_{0}^{t} e^{-\alpha(t-\tau)} g(\tau) d \tau, \quad \forall t \geq 0
    		\end{displaymath}
    		for any finite constant $\alpha$.  	
    }
    \end{lemmax}

	\begin{theoremx}\emph{
		Under Assumptions 1, 2, 4, and 5, given scalars $\mu_1, \mu_2, \sigma>0$, if there exist symmetric positive definite matrices $\bm{P}\in \mathbb{R} ^ {m \times m}$, $\bm{G}_1\in \mathbb{R} ^ {m \times m}$, $\bm{G}_2\in \mathbb{R} ^ {m \times m}$, matrices $\bm{X}\in \mathbb{R} ^ {m \times n_y}$, $\bm{F}\in \mathbb{R} ^ {m \times n_y}$, and a positive constant $\varepsilon_1$ such that 
		the following linear matrix inequality (LMI) is satisfied:
		\begin{equation} \label{e15}
		\begin{aligned}
		\bm{\Xi} \mathop {=}\limits^\Delta  \left[ {\begin{array}{*{20}{c}}
			{\bm{\Xi}_{11}} & \;\; *  & \;\; *   \\
			\frac{1}{\sigma}(\bm{X}\bm{C}_s-\bm{P}\bm{A}_s) & \;\; {\bm{\Xi}_{22}}&  \;\;*   \\
			{{\bm{X}^T}} & \;\; \bm{F}^T-\frac{1}{\sigma}{{\bm{X}^T}}& \;\;\;\;\;\;-\varepsilon_1\bm{I} \\	
			\end{array} } \right]< 0,
		\end{aligned}
		\end{equation} 
		where 
		\begin{displaymath}
		\bm{\Xi}_{11}\mathop {=}\limits^\Delta  \bm{P}{{\bm{A}}_s} + {\bm{A}}_s^T\bm{P} - \bm{X}\bm{C}_s - \bm{C}_s^T{\bm{X}^T},
		\end{displaymath}
		\begin{displaymath}
		{\bm{\Xi}_{22}}\mathop {=}\limits^\Delta  -\frac{2}{\sigma}\bm{P}+\frac{1}{\sigma \mu_1}\bm{G}_1+\frac{1}{\sigma\mu_2}\bm{G}_2,
		\end{displaymath}
		\begin{equation}\label{e16}
		\bm{X}=\bm{PL},
		\end{equation}
		and the following condition holds:
		\begin{equation}\label{e16+}
		\bm{P}=\bm{FC}_s,
		\end{equation}
		then the adaptive source estimation algorithm in (\ref{e17}) can realize $\bm{e}_x(t)$ and $\bm{e}_f(t)$ uniformly ultimately bounded (UUB)} where the symmetric positive definite matrix $\bm{\Gamma}\in \mathbb{R} ^ {m \times m}$ denotes the learning rate.
	\end{theoremx}
	
	\textbf{Proof}.
	Choose the Lyapunov candidate as follows:
	\begin{equation}\label{e18}
	V(t)=\bm{e}^T_x(t)\bm{Pe}_x(t)+\frac{1}{\sigma}\bm{e}^T_f(t)\bm{\Gamma}^{-1}\bm{e}_f(t).
	\end{equation}
	Combing (\ref{e13}), (\ref{e17}), (\ref{e16}), and (\ref{e16+}), its derivative with respect to time is: 
	\begin{equation}\label{e20}
	\begin{aligned}
	\dot {V}(t) &= \bm{e}_x^T(t)(\bm{P}{{\bm{A}}_s} + {\bm{A}}_s^T\bm{P} - \bm{X}{\bm{C}_s} - \bm{C}_s^T{\bm{X}^T}){\bm{e}_x}(t)\\ 
	&+ \frac{2}{\sigma}\bm{e}_f^T(t)(\bm{X}{\bm{C}_s}-\bm{PA}_s){\bm{e}_x}(t)+ 2\bm{y}_f(t)^T{\bm{X}^T}{\bm{e}_x}(t)\\
	&+ 2\bm{y}_f^T(t)(\bm{F}^T-\frac{1}{\sigma}\bm{X}^T){\bm{e}_f}(t)- \frac{2}{\sigma}\bm{e}_f^T(t){\bm{\Gamma} ^{ - 1}}{\dot {\bm{f}}}_s(t)\\
	&-\frac{2}{\sigma}\bm{e}_f^T(t)\bm{Pe}_f(t)+\frac{2}{\sigma}\bm{e}_f^T(t)\bm{F}\dot{\bm{y}}_f(t).
	\end{aligned}
	\end{equation} 
	Combing Lemma 1, Assumption 1, and Assumption 2, it can be obtained that: 
	\begin{equation}\label{e21}
	\begin{aligned}
	- \frac{2}{\sigma}\bm{e}_f^T(t){\bm{\Gamma} ^{ - 1}}{{\dot {\bm{f}}}_s}(t) &\leqslant \frac{1}
	{\sigma \mu_1 }\bm{e}_f^T(t)\bm{G}_1\bm{e}_f(t) \\
	&+ \frac{\mu_1}{\sigma} \dot {\bm{f}}_s^T(t){\bm{\Gamma} ^{ - 1}}\bm{G}^{-1}_1{\bm{\Gamma} ^{ - 1}}{{\dot {\bm{f}}}_s}(t) \hfill \\
	&\leqslant \frac{1}{\sigma \mu_1  }\bm{e}_f^T(t)\bm{G}_1\bm{e}_f(t)\\
	&+ \frac{\mu_1}{\sigma} f_1{\lambda _{\max }}({\bm{\Gamma} ^{ - 1}}\bm{G}^{-1}_1{\bm{\Gamma} ^{ - 1}}). \hfill \\ 
	\end{aligned} 
	\end{equation}
	
	\begin{equation}\label{e21-}
	\begin{aligned}
	\frac{2}{\sigma}\bm{e}_f^T(t)\bm{F}\dot{\bm{y}}_f(t) &\leqslant \frac{1}
	{\sigma\mu_2 }\bm{e}_f^T(t)\bm{G}_2\bm{e}_f(t)\\
	&+ \frac{\mu_2}{\sigma}\dot {\bm{y}}_f^T(t)\bm{F}^T\bm{G}_2^{-1}\bm{F}{{{\dot{\bm{y}}}}_f}(t) \hfill \\
	&\leqslant \frac{1}{\sigma\mu_2 }\bm{e}_f^T(t)\bm{G}_2\bm{e}_f(t)\\ 
	&+ \frac{\mu_2}{\sigma}{\lambda _{\max }}(\bm{F}^T\bm{G}_2^{-1}\bm{F}){\left\| {{\dot{\bm{y}}_f}(t)} \right\|_{peak}^2}. \hfill \\ 
	\end{aligned} 
	\end{equation}
	Substituting (\ref{e21}) and (\ref{e21-}) into (\ref{e20}) and considering Assumption 1 yields:
	\begin{equation} \label{e22}
	\dot {V}(t) \leqslant {\bm{\xi} ^T}(t)\bm{\Xi} \bm{\xi} (t)+\beta+{\varepsilon _1}{\left\| {{{\bm{y}}_f}(t)} \right\|_{peak}^2}+\varepsilon_2{\left\| {{\dot{\bm{y}}_f}(t)} \right\|_{peak}^2}, 
	\end{equation}
	where 
	\begin{equation*} 
	\begin{aligned}
	\bm{\xi} (t) &\mathop {=}\limits^\Delta   \left[ \begin{gathered}
	{\bm{e}_x}(t)  \hfill \\
	{\bm{e}_f}(t)  \hfill \\ 
	{\bm{y}_f}(t)  \hfill \\
	\end{gathered}  \right], \\
	\bm{\Xi} &\mathop {=}\limits^\Delta  \left[ {\begin{array}{*{20}{c}}
		{\bm{\Xi}_{11}} & \;\; *  & \;\; *   \\
		\frac{1}{\sigma}(\bm{X}\bm{C}_s-\bm{P}\bm{A}_s) & \;\; {\bm{\Xi}_{22}}&  \;\;*   \\
		{{\bm{X}^T}} & \;\; \bm{F}^T-\frac{1}{\sigma}{{\bm{X}^T}}& \;\;\;\;\;\;-\varepsilon_1\bm{I} \\	
		\end{array} } \right],\\
	\bm{\Xi}_{11}&\mathop {=}\limits^\Delta  \bm{P}{{\bm{A}}_s} + {\bm{A}}_s^T\bm{P} - \bm{X}\bm{C}_s - \bm{C}_s^T{\bm{X}^T},\\
	{\bm{\Xi}_{22}}&\mathop {=}\limits^\Delta  -\frac{2}{\sigma}\bm{P}+\frac{1}{\sigma \mu_1}\bm{G}_1+\frac{1}{\sigma\mu_2}\bm{G}_2,\\  
	\beta&\mathop {=}\limits^\Delta  \frac{\mu_1}{\sigma} f_1{\lambda _{\max }}({\bm{\Gamma} ^{ - 1}}\bm{G}^{-1}_1{\bm{\Gamma} ^{ - 1}}), \varepsilon_2\mathop {=}\limits^\Delta  \frac{\mu_2}{\sigma}{\lambda _{\max }}(\bm{F}^T\bm{G}_2^{-1}\bm{F}).	
	\end{aligned}
	\end{equation*}

	
	Hence, when $\bm{\Xi}<0$, one can obtain that: 
	\begin{equation}
	\label{e23}
	\begin{aligned}
	\dot V(t) &\leqslant  - {\lambda _{\min }}( - \bm{\Xi} ){\left\| {\bm{\xi} (t)} \right\|^2} + \beta+{\varepsilon _1}{\left\| {{{\bm{y}}_f}(t)} \right\|_{peak}^2}\\
	&+\varepsilon_2{\left\| {{\dot{\bm{y}}_f}(t)} \right\|_{peak}^2} \\
	&= - {\lambda _{\min }}( - \bm{\Xi} )({\left\| {{\bm{e}_x}(t)} \right\|^2} + {\left\| {{\bm{e}_f}(t)} \right\|^2}+{\left\| {{\bm{y}_f}(t)} \right\|^2})\\
	&+\beta+{\varepsilon _1}{\left\| {{{\bm{y}}_f}(t)} \right\|_{peak}^2}+\varepsilon_2{\left\| {{\dot{\bm{y}}_f}(t)} \right\|_{peak}^2}\\
	&\leqslant- {\lambda _{\min }}( - \bm{\Xi} )({\left\| {{\bm{e}_x}(t)} \right\|^2} + {\left\| {{\bm{e}_f}(t)} \right\|^2})\\
	& +\beta+{\varepsilon _1}{\left\| {{{\bm{y}}_f}(t)} \right\|_{peak}^2}+\varepsilon_2{\left\| {{\dot{\bm{y}}_f}(t)} \right\|_{peak}^2}. 
	\end{aligned}
	\end{equation}
	
	According to the definition of $V(t)$ in (\ref{e18}), it can be derived that:
	\begin{equation}
	\label{e24}
	\begin{aligned}
	{V}(t) &\leqslant {\lambda _{\max }}(\bm{P}){\left\| {{\bm{e}_x}(t)} \right\|^2} + \frac{1}{\sigma}{\lambda _{\max }}({\bm{\Gamma} ^{ - 1}}){\left\| {{\bm{e}_f}(t)} \right\|^2} \\ 
	&\leqslant \max \{ {\lambda _{\max }}(\bm{P}), \frac{1}{\sigma}{\lambda _{\max }}({\bm{\Gamma} ^{ - 1}})\} ({\left\| {{\bm{e}_x}(t)} \right\|^2} + {\left\| {{\bm{e}_f}(t)} \right\|^2}). \\ 
	\end{aligned}
	\end{equation}
	
	Combing (\ref{e23}) and (\ref{e24}), it can be obtained that:
	\begin{equation}
	\label{e25}
	\dot{V}(t)\leqslant-\alpha V(t)+\beta+{\varepsilon _1}{\left\| {{{\bm{y}}_f}(t)} \right\|_{peak}^2}+\varepsilon_2{\left\| {{\dot{\bm{y}}_f}(t)} \right\|_{peak}^2},
	\end{equation}
	where 
	\begin{displaymath}
	\alpha=\frac{{\lambda _{\min }}( - \bm{\Xi} )}{\max \{ {\lambda _{\max }}(\bm{P}),  \frac{1}{\sigma}{\lambda _{\max }}({\bm{\Gamma} ^{ - 1}})\}}.
	\end{displaymath}
	
	By Lemma 2, it can be obtained that:
	\begin{equation}
	\label{e25+}
	\begin{aligned}
	{V}(t)&\leqslant e^{-\alpha t}V(0)+(\beta+{\varepsilon _1}{\left\| {{{\bm{y}}_f}(t)} \right\|_{peak}^2}+\varepsilon_2{\left\| {{\dot{\bm{y}}_f}(t)} \right\|_{peak}^2})\\
	&\int_{0}^{t} e^{-\alpha(t-\tau)} d \tau\\
	&=e^{-\alpha t}V(0)+(\beta+{\varepsilon _1}{\left\| {{{\bm{y}}_f}(t)} \right\|_{peak}^2}+\varepsilon_2{\left\| {{\dot{\bm{y}}_f}(t)} \right\|_{peak}^2})\\
	&\frac{1}{\alpha}\left(1-e^{-\alpha t}\right)\\
	&\leqslant e^{-\alpha t}V(0)+(\frac{\beta}{\alpha}+\frac{\varepsilon _1}{\alpha}{\left\| {{{\bm{y}}_f}(t)} \right\|_{peak}^2}+\frac{\varepsilon_2}{\alpha}{\left\| {{\dot{\bm{y}}_f}(t)} \right\|_{peak}^2})\\
	&\sup _{t \in[0, \infty)}\left\{1-e^{-\alpha t}\right\}\\
	&\leqslant e^{-\alpha t}V(0)+(\frac{\beta}{\alpha}+\frac{\varepsilon _1}{\alpha}{\left\| {{{\bm{y}}_f}(t)} \right\|_{peak}^2}+\frac{\varepsilon_2}{\alpha}{\left\| {{\dot{\bm{y}}_f}(t)} \right\|_{peak}^2})\\
	&\leqslant e^{-\alpha t}V(0)+(\sqrt{\frac{\beta}{\alpha}}+\sqrt{\frac{\varepsilon _1}{\alpha}}{\left\| {{{\bm{y}}_f}(t)} \right\|_{peak}}\\
	&+\sqrt{\frac{\varepsilon_2}{\alpha}}{\left\| {{\dot{\bm{y}}_f}(t)} \right\|_{peak}})^2\\
	\end{aligned}
	\end{equation}

	Meanwhile, considering (\ref{e18}), it can be derived that:
	\begin{equation}
	\label{e26-}
	\begin{aligned}
	{V}(t) &\ge {\lambda _{\min }}(\bm{P}){\left\| {{\bm{e}_x}(t)} \right\|^2} +  \frac{1}{\sigma}{\lambda _{\min }}({\bm{\Gamma} ^{ - 1}}){\left\| {{\bm{e}_f}(t)} \right\|^2} \\ 
	&\ge \min \{ {\lambda _{\min }}(\bm{P}),  \frac{1}{\sigma}{\lambda _{\min }}({\bm{\Gamma} ^{ - 1}})\} ({\left\| {{\bm{e}_x}(t)} \right\|^2} + {\left\| {{\bm{e}_f}(t)} \right\|^2}). \\ 
	\end{aligned}
	\end{equation} 
	Combining (\ref{e25+}) with (\ref{e26-}), it can be derived that:

   \begin{equation}
   	\label{e25++}
   	\begin{aligned}
   	 {\left\| {{\bm{e}_x}(t)} \right\|^2} + {\left\| {{\bm{e}_f}(t)} \right\|^2}&\leqslant \frac{e^{-\alpha t}V(0)}{\min \{ {\lambda _{\min }}(\bm{P}),  \frac{1}{\sigma}{\lambda _{\min }}({\bm{\Gamma} ^{ - 1}})\}}+\rho^2\\
   	 &\leqslant (\sqrt{\frac{e^{-\alpha t}V(0)}{\min \{ {\lambda _{\min }}(\bm{P}),  \frac{1}{\sigma}{\lambda _{\min }}({\bm{\Gamma} ^{ - 1}})\}}}+\rho)^2
   	\end{aligned}
   \end{equation}
   where 
   \begin{equation}
   \begin{aligned}
   \label{e25+++}
   \rho&=\sqrt{\frac{1}{\min \{ {\lambda _{\min }}(\bm{P}),  \frac{1}{\sigma}{\lambda _{\min }}({\bm{\Gamma} ^{ - 1}})\}}}(\sqrt{\frac{\beta}{\alpha}}+\sqrt{\frac{\varepsilon _1}{\alpha}}{\left\| {{{\bm{y}}_f}(t)} \right\|_{peak}}\\
   &+\sqrt{\frac{\varepsilon_2}{\alpha}}{\left\| {{\dot{\bm{y}}_f}(t)} \right\|_{peak}}).
   \end{aligned}
   \end{equation}
   Hence it can be further derived that
   \begin{equation}
   \label{e25++++}
   \begin{aligned}
   {\left\| {{\bm{e}_x}(t)} \right\|}&\leqslant \sqrt{\frac{e^{-\alpha t}V(0)}{\min \{ {\lambda _{\min }}(\bm{P}),  \frac{1}{\sigma}{\lambda _{\min }}({\bm{\Gamma} ^{ - 1}})\}}}+\rho,\\
   {\left\| {{\bm{e}_f}(t)} \right\|}&\leqslant \sqrt{\frac{e^{-\alpha t}V(0)}{\min \{ {\lambda _{\min }}(\bm{P}),  \frac{1}{\sigma}{\lambda _{\min }}({\bm{\Gamma} ^{ - 1}})\}}}+\rho.
   \end{aligned}
   \end{equation}
   Recall Assumption 1, it can be obtained that $0\leqslant\rho<\infty$. Hence (\ref{e25++++}) implies that there exists a $T \to \infty$ such that ${\left\| {{\bm{e}_x}(t)} \right\|}\leqslant \rho, {\left\| {{\bm{e}_f}(t)} \right\|}\leqslant \rho$, for all $t>T$. That is to say, $\bm{e}_x(t)$ and $\bm{e}_f(t)$ are UUB with ultimate bound $\rho$.  This completes the proof.\;\;$\blacksquare$


	\begin{corollaryx}\emph{
		The abnormal S-T source estimation error of the original PDE system in (\ref{e1})-(\ref{e3}) can be obtained as
		\begin{equation*}
		{e_f}(z,t) = \hat f(z,t) - f(z,t),
		\end{equation*}	 
		where $\hat{f}(z,t)=\bm{\phi} _s^T(z){\hat{\bm{f}}_s}(t)$, as shown in (\ref{eqn:t/ssynthesis}). Hence the square of the norm for the abnormal S-T source estimation error in $\mathcal{H}$ can be obtained as
		\begin{equation*}
		\begin{aligned}
		\left\| {{e_f}(z,t)} \right\|_2^2 &= \left\| {\hat f(z,t) - f(z,t)} \right\|_2^2 \hfill \\
		&= \left\| {\bm{\phi} _s^T(z){{\hat {\bm{f}}}_s}(t) - \bm{\phi} _s^T(z){\bm{f}_s}(t) - \bm{\phi} _f^T(z){\bm{f}_f}(t)} \right\|_2^2 \hfill \\
		&= \left\| {\bm{\phi} _s^T(z){\bm{e}_f}(t) - \bm{\phi} _f^T(z){\bm{f}_f}(t)} \right\|_2^2 \hfill \\
		&= {\left\| {{\bm{e}_f}(t)} \right\|^2} + {\left\| {{\bm{f}_f}(t)} \right\|^2}. \hfill \\ 
		\end{aligned}
		\end{equation*} 
		Combing with Theorem 1 and Assumption 3, it can be concluded that the abnormal S-T source estimation error ${e_f}(z,t)$ is UUB if the conditions in Theorem 1 are satisfied, with ultimate bound $\rho+\sqrt{f_2}$.}  
	\end{corollaryx}

	\begin{remark}\emph{
		The inequality (\ref{e15}) can be solved by the MATLAB LMI toolbox. However, the equality condition in (\ref{e16+}) makes the problem difficult to solve. Therefore, this condition is transformed into the following problem: Minimize $\eta$ subject to (\ref{e15}), (\ref{e16}) and} 
		\begin{equation}
		\label{e26+}
		\left[ {\begin{array}{*{20}{c}}
			{\eta \bm{I}} &  *   \\
			{{{(\bm{P} - \bm{F}{\bm{C}_s})}^T}} & {\eta \bm{I}}  \\
			
			\end{array} } \right] > 0.
		\end{equation}  
	\end{remark}
	
	%

	
	\section{Numerical Simulation}\label{sec:Numerical Simulation}
	Consider a thin rod whose temperature distribution can be modeled by the following parabolic PDE:
	\begin{equation}\label{e26}
	\begin{aligned}
	\frac{{\partial x(z,t)}}
	{{\partial t}}= \frac{{{\partial ^2}x(z,t)}}
	{{\partial {z^2}}} + {\beta _U}(b_u(z)u(t) - x(z,t))+ f(z,t),
	\end{aligned}
	\end{equation}  
	\begin{equation} \label{e27}
	\bm{y}(t) = \left[ \begin{gathered}
	{\int_0^\pi  {\delta (z - \frac{\pi }{4})x(z,t)dz} }   \hfill \\
	{\int_0^\pi  {\delta (z - \frac{3\pi }{4})x(z,t)dz} }   \hfill \\ 
	\end{gathered}  \right]   = \left[ \begin{gathered}
	{x(\frac{\pi }{4},t)}   \hfill \\
	{x(\frac{3\pi }{4},t)}   \hfill \\ 
	\end{gathered}  \right],
	\end{equation}
	subject to the Dirichlet boundary conditions:
	\begin{equation}
	x(0,t) = 0,\;x(\pi ,t) = 0,
	\end{equation}
	where $x(z,t)$ denotes the dimensionless temperature of the rod; $\beta_U$ denotes a dimensionless heat transfer coefficient; $u(t)$ is the manipulated input (temperature of the cooling medium); $\bm{y}(t)$ are the thermocouple measurements at point $z =\pi/{4}$ and $z =3\pi/{4}$. The typical value of $\beta_U$ is given as $2$.
	The actuator distribution function is set as:
	\begin{displaymath}
	{b_u}(z) = \sqrt {\frac{2}{\pi }}\sin (z).
	\end{displaymath}
	And the control input is selected as: 
	\begin{displaymath}
	u(t)=1.
	\end{displaymath} 
	The eigenvalue problem for the spatial differential operator:
	\begin{equation*}
	\begin{aligned}
	{\mathcal{A}}x &= \frac{{{\partial ^2}x}}{{\partial {z^2}}}-\beta_Ux,\;\\
	x \in \mathcal{S}(\mathcal{A}) &= \{ x \in \mathcal{L}_2([{0},{\pi}];\mathbb{R}); x(0,t) = 0,x(\pi ,t) = 0\} 
	\end{aligned}
	\end{equation*}
	can be directly solved as:
	\begin{equation*}
	{\lambda _j} =  - {j^2}-2,\;{\phi _j}(z) = \sqrt {\frac{2}
		{\pi }} \sin (jz),\;j = 1,\cdots,\infty. 
	\end{equation*}
	Details of solving procedures for this kind of eigenvalue problem can be referred to Cheaper 4 of~\cite{ray1989control}. The first two eigenvalues are considered as the dominant ones, thus $\varepsilon  = \left| {{\lambda _1}} \right|/\left| {{\lambda _{3}}} \right| \approx 0.273$. Using the procedures discussed in Section~\ref{sec:Adaptive fault diagnosis observer design}, the following $2$-dimensional slow subsystem is derived: 
	\begin{equation} \label{e29}
	\begin{aligned}
	{{\dot {\bm{x}}}_s}(t) &= {{\bm{A}}_s}{\bm{x}_s}(t)+ {\bm{B}_{u,s}}\bm{u}(t)+\bm{f}_s(t), \hfill \\
	\bm{y}_s(t) &= {\bm{C}_{s}{\bm{x}_s}(t)}, \hfill \\
	\end{aligned}
	\end{equation}
	where 
	\begin{equation*}
	\bm{x}_s(t)=\left[ \begin{gathered}
	x_1(t)  \hfill \\
	x_2(t)  \hfill \\ 
	\end{gathered}  \right],\\
	{\bm{A}}_s=\left[ {\begin{array}{*{20}{c}}
		{ - 3} & 0  \\
		0 & { - 6}  \\	
		\end{array} } \right],
	{\bm{B}_{u,s}}=\left[ \begin{gathered}
	2  \hfill \\
	0  \hfill \\ 
	\end{gathered}  \right],\\
	\end{equation*}
	and 
	\begin{equation*}
	\bm{C}_s=\int_0^\pi  \left[ \begin{gathered}
	{\delta (z - \frac{\pi }{4})}   \hfill \\
	{\delta (z - \frac{3\pi }{4})}   \hfill \\ 
	\end{gathered}  \right] \left[\phi_1(z)\;\phi_2(z)\right] dz =\left[ {\begin{array}{*{20}{c}}
		\sqrt {\frac{1}{\pi}} & \sqrt {\frac{2}{\pi}}  \\
		\sqrt {\frac{1}{\pi}} & {-\sqrt {\frac{2}{\pi}}}  \\	
		\end{array} } \right].
	\end{equation*}
	It can be found that Assumption 4 is satisfied. In this numerical simulation, consider the following abnormal S-T source according to Remark 4:  
	\begin{displaymath}
	f(z,t)=[\phi_1(z)\;\phi_2(z)]\left[ \begin{gathered}
	{f}_{1}(t)  \hfill \\
	{f}_{2}(t)  \hfill \\ 
	\end{gathered}  \right].
	\end{displaymath}
	Abnormal source $f_1(t)$ can be considered as the actuator fault due to the fact that $b_u(z)=\phi_1(z)$ is set.
	
	To evaluate the performance of the proposed abnormal S-T source identification method, the following kinds of abnormal sources are considered:
	\begin{equation*}
	\begin{aligned}
	f_1(t) &= \left\{ \begin{gathered}
	0,\;0 \leqslant t < 10(\sec ) \hfill \\
	2,\; 10 \leqslant t \leqslant 80(\sec ) \hfill \\ 
	\end{gathered}  \right. \hfill \\ 
	f_2(t) &= \left\{ \begin{gathered}
	0,\;0 \leqslant t < 40(\sec ) \hfill \\
	3,\; 40 \leqslant t \leqslant 80(\sec ) \hfill \\ 
	\end{gathered}  \right. \hfill \\ 
	\end{aligned}
	\end{equation*}   
	and 
	\begin{equation*}
	\begin{aligned}
	f_1(t) &= \left\{ \begin{gathered}
	0,\;\;\;\;\;\;\;\;\;\;\;\;\;\;\;\;\;\;\;\;\;\;\;\;\;0 \leqslant t < 10(\sec ) \hfill \\
	2-e^{-0.01(t-10)},\; 10 \leqslant t \leqslant 80(\sec ) \hfill \\ 
	\end{gathered}  \right. \hfill \\ 
	f_2(t) &= \left\{ \begin{gathered}
	0,\;\;\;\;\;\;\;\;\;\;\;\;\;\;\;\;\;\;\;\;\;\;\;\;\;0 \leqslant t < 40(\sec ) \hfill \\
	3-e^{-0.02(t-40)},\; 40 \leqslant t \leqslant 80(\sec ) \hfill \\  
	\end{gathered}  \right. \hfill \\ 
	\end{aligned}
	\end{equation*}  
	The first kind of abnormal source can be considered as an abrupt source while the second is an incipient one. Based on the definition of $\bm{f}_s(t)$ in (\ref{e8}), it can be obtained that: 
	\begin{displaymath}
	\bm{f}_s(t)=\left[ \begin{gathered}
	{f}_{s1}(t)  \hfill \\
	{f}_{s2}(t)  \hfill \\ 
	\end{gathered}  \right]
	=\left[ \begin{gathered}
	{f}_1(t)  \hfill \\
	{f}_2(t)  \hfill \\ 
	\end{gathered}  \right].
	\end{displaymath}   
	
	According to Theorem 1, choosing $\mu_1=1, \mu_2=1$, and $\sigma=1$ and solving (\ref{e15}), (\ref{e16}), and (\ref{e26+}), one can obtain that: 
	\begin{equation*}
	\begin{aligned}
	\eta&=9.4277\times 10^{-12},\\
	\bm{P}&=\left[ {\begin{array}{*{20}{c}}
		{0.1774} & 0  \\
		0 & {0.0609}  \\	
		\end{array} } \right],
	\bm{G}_1=\left[ {\begin{array}{*{20}{c}}
		{0.0193} & 0  \\
		0 & {0.0102}  \\	
		\end{array} } \right],\\
	\bm{G}_2&=\left[ {\begin{array}{*{20}{c}}
		{0.0193} & 0  \\
		0 & {0.0102}  \\	
		\end{array} } \right],
	\bm{X}=\left[ {\begin{array}{*{20}{c}}
		{-0.1106}&-0.1106   \\
		-0.1588&0.1588   \\	
		\end{array} } \right],\\
	\bm{F}&=\left[ {\begin{array}{*{20}{c}}
		{0.1572}&0.1572   \\
		0.0382&-0.0382   \\	
		\end{array} } \right],
	\bm{L}=\left[ {\begin{array}{*{20}{c}}
		{-0.6231}&-0.6231   \\
		-2.6069&2.6069   \\	
		\end{array} } \right],
	\end{aligned}
	\end{equation*}  
	by the MATLAB LMI toolbox.  
	
	Moreover, the learning rate is chosen as $\bm{\Gamma}=diag(100,100)$.  
	The original PDE system in (\ref{e26}) is solved numerically by the finite difference method (FDM) \cite{strikwerda2004finite} with the sampling time $\Delta t=0.01\sec$. By using the adaptive state observer (\ref{e12}) and the adaptive source estimation algorithm (\ref{e17}), the source estimation of the slow subsystem $\hat{\bm{f}}_s(t)=[\hat{f}_{s1}(t)\;\hat{f}_{s2}(t)]^T$ can be gained and the abnormal S-T source estimation for the original PDE system in (\ref{e26}) can be obtained by:
	\begin{displaymath}
	\hat{f}(z,t)=[\phi_1(z)\;\phi_2(z)]\left[ \begin{gathered}
	\hat{f}_{s1}(t)  \hfill \\
	\hat{f}_{s2}(t)  \hfill \\ 
	\end{gathered}  \right],
	\end{displaymath}    
	as shown in (\ref{eqn:t/ssynthesis}).
	
	\begin{figure}[!h]
		\centering
		\subfigure[The first output of the original PDE system $y_1(t)$, the slow subsystem $y_{s1}(t)$, and its estimation $\hat{y}_{s1}(t)$]{
			\includegraphics[width=3.8cm]{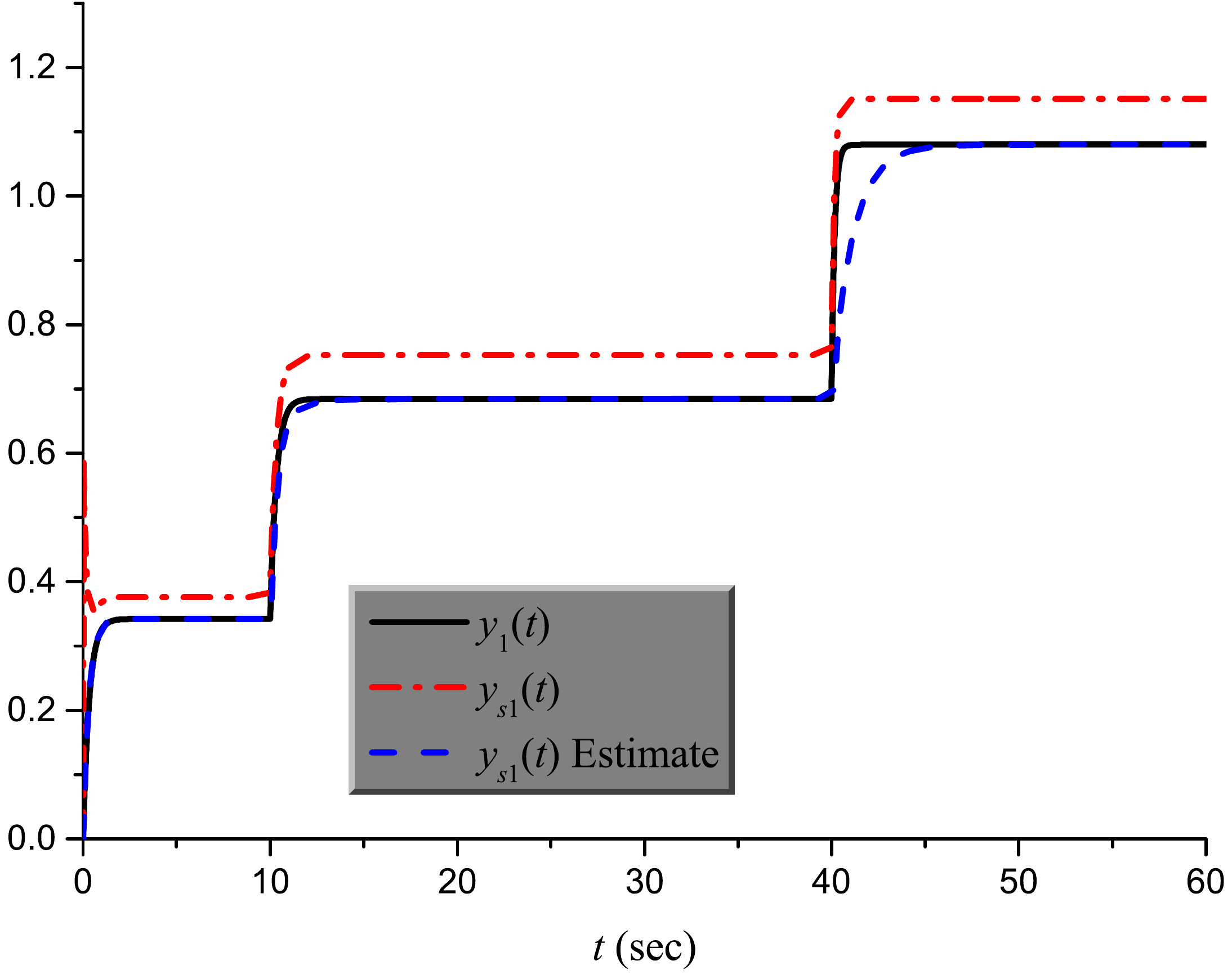}	
			\label{fig:2a}	
		}
		\subfigure[The second output of the original PDE system $y_2(t)$, the slow subsystem $y_{s2}(t)$, and its estimation $\hat{y}_{s2}(t)$]{
			\includegraphics[width=3.8cm]{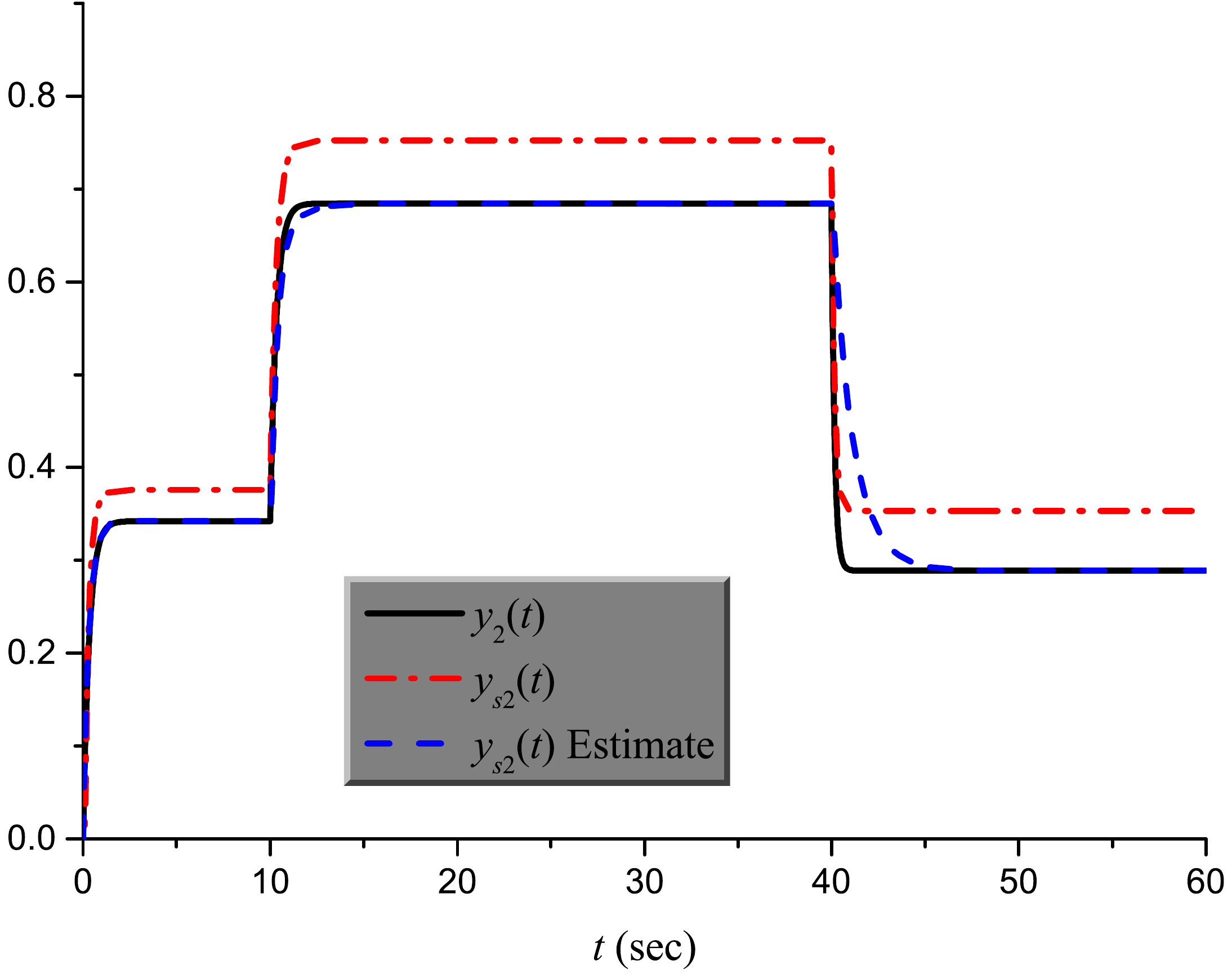}	
			\label{fig:2b}	
		}
		\subfigure[$\bm{f}_s(t)$ and its estimation $\hat{\bm{f}}_s(t)$]{
			\includegraphics[width=3.8cm]{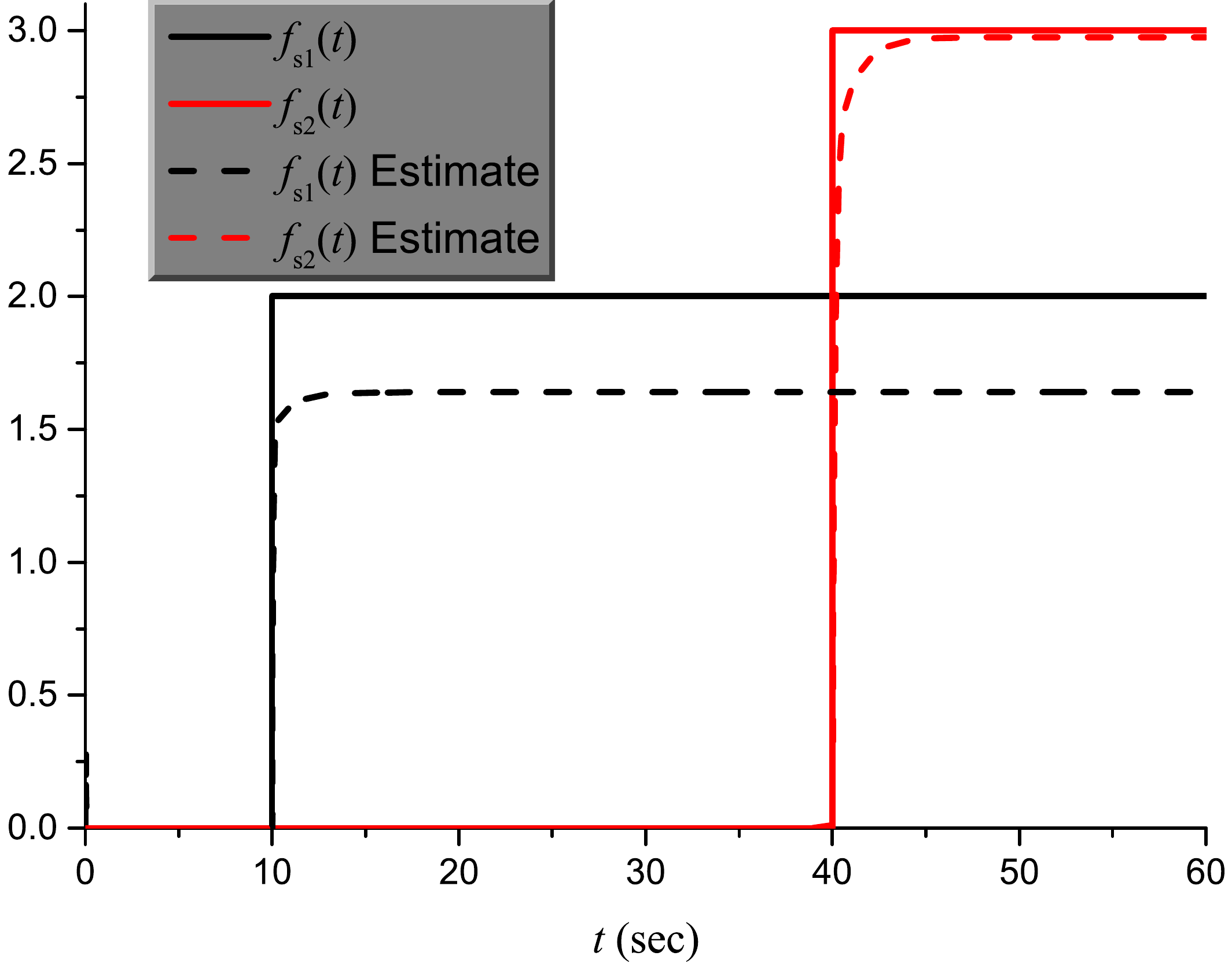}	
			\label{fig:2c}}	

		\caption{Estimation results of abnormal source 1. (Abrupt source)}
	\end{figure}

    \begin{figure}[!h]
    	\centering
    	\includegraphics[width=7cm]{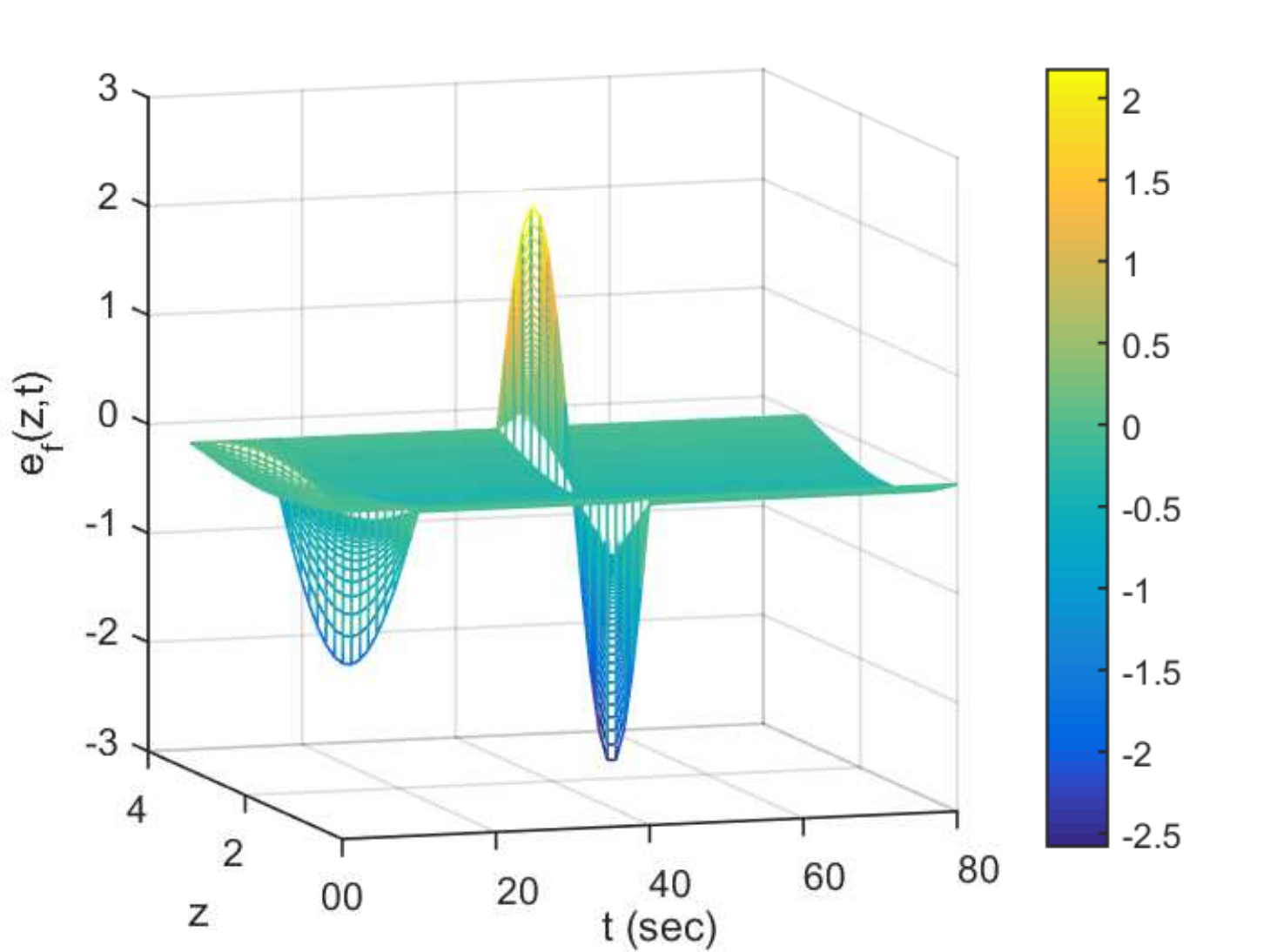}
    	\caption{$e_f(z,t)$ of abnormal source 1. (Abrupt source).}
    	\label{fig:ef1}
    \end{figure}

	\begin{figure}[!h]
		\centering
		\subfigure[The first output of the original PDE system $y_1(t)$, the slow subsystem $y_{s1}(t)$, and its estimation $\hat{y}_{s1}(t)$]{
			\includegraphics[width=3.8cm]{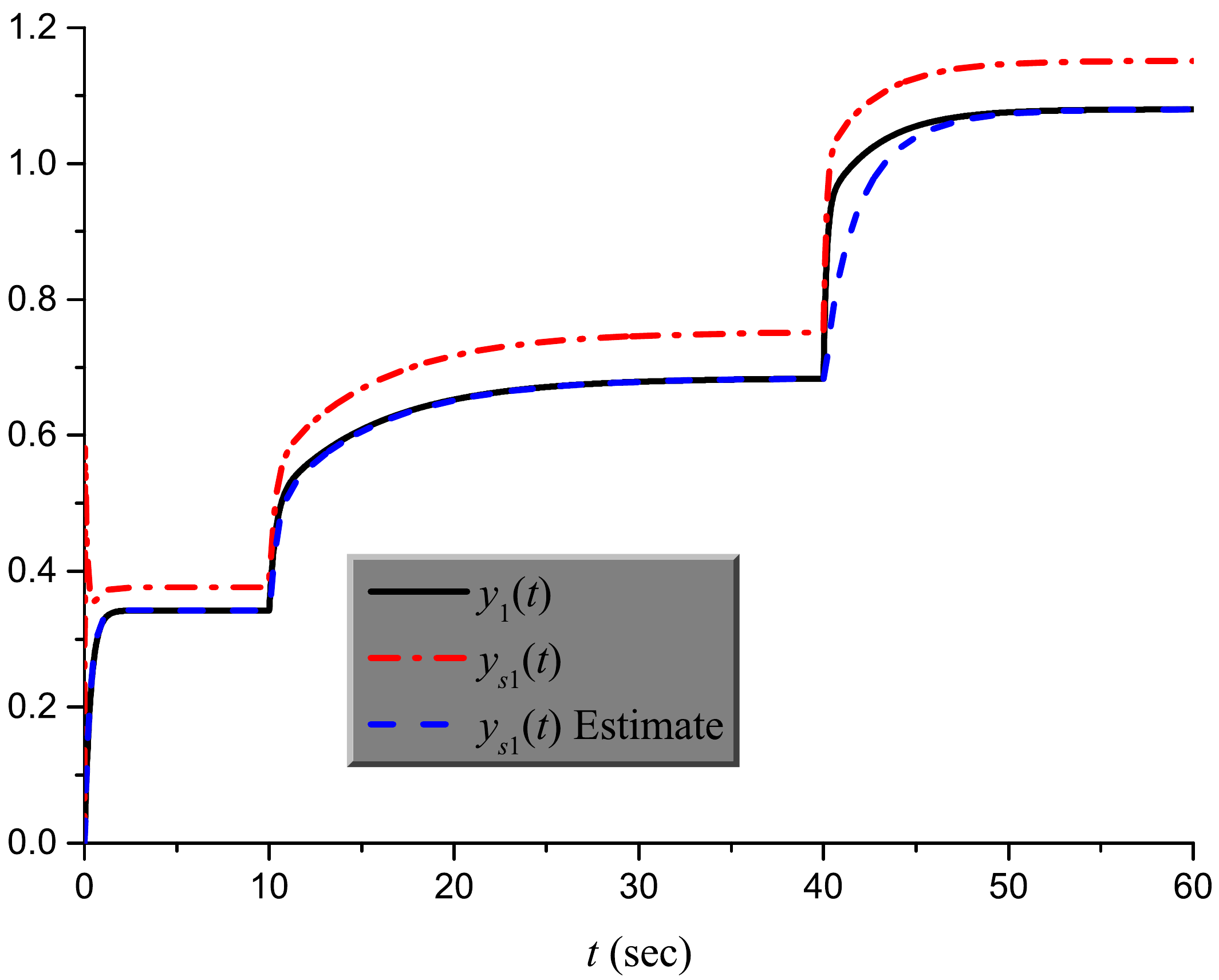}
			\label{fig:4a}		
		}
		\subfigure[The second output of the original PDE system $y_2(t)$, the slow subsystem $y_{s2}(t)$, and its estimation $\hat{y}_{s2}(t)$]{
			\includegraphics[width=3.8cm]{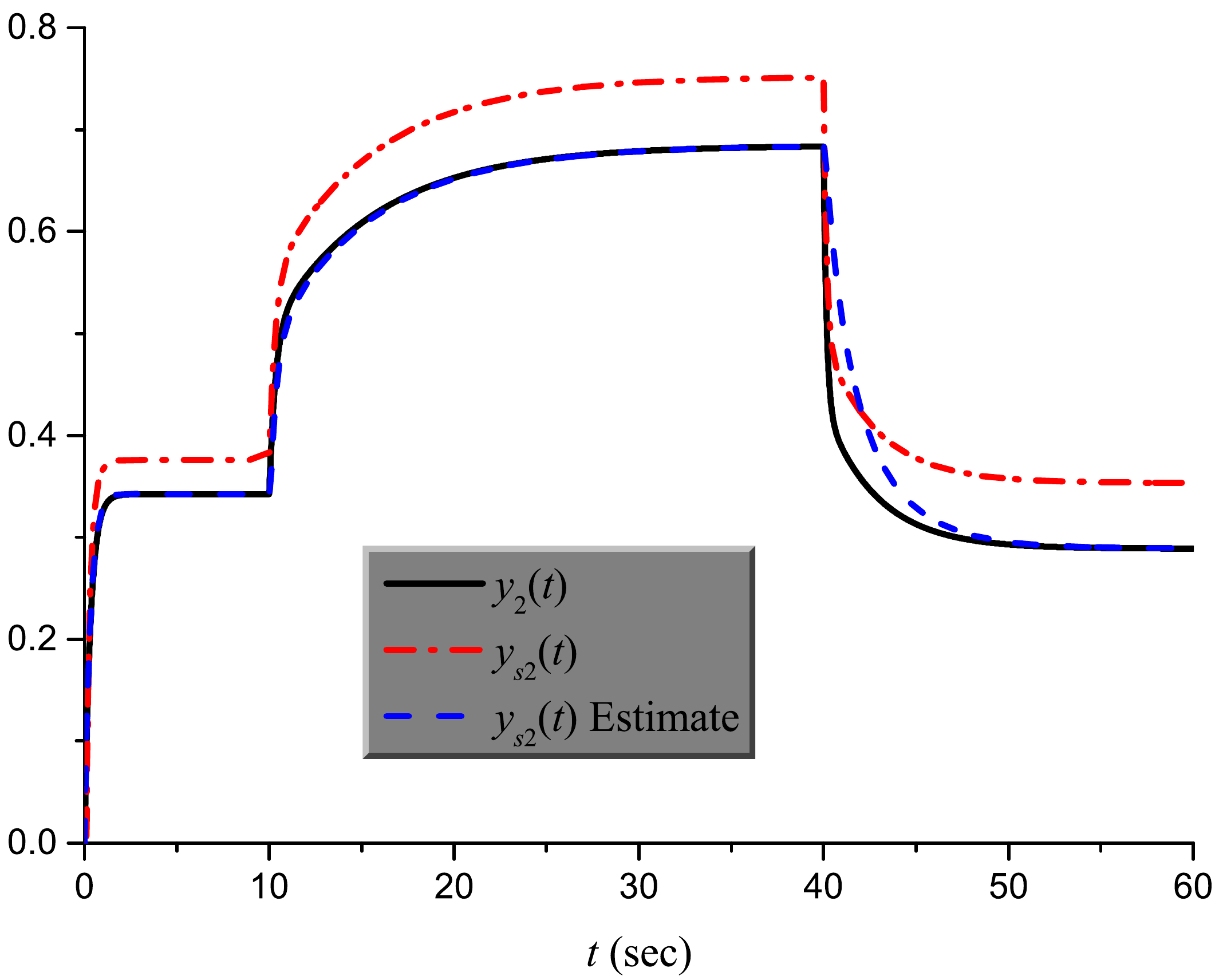}	
			\label{fig:4b}	
		}
		\subfigure[$\bm{f}_s(t)$ and its estimation $\hat{\bm{f}}_s(t)$]{
				\includegraphics[width=3.8cm]{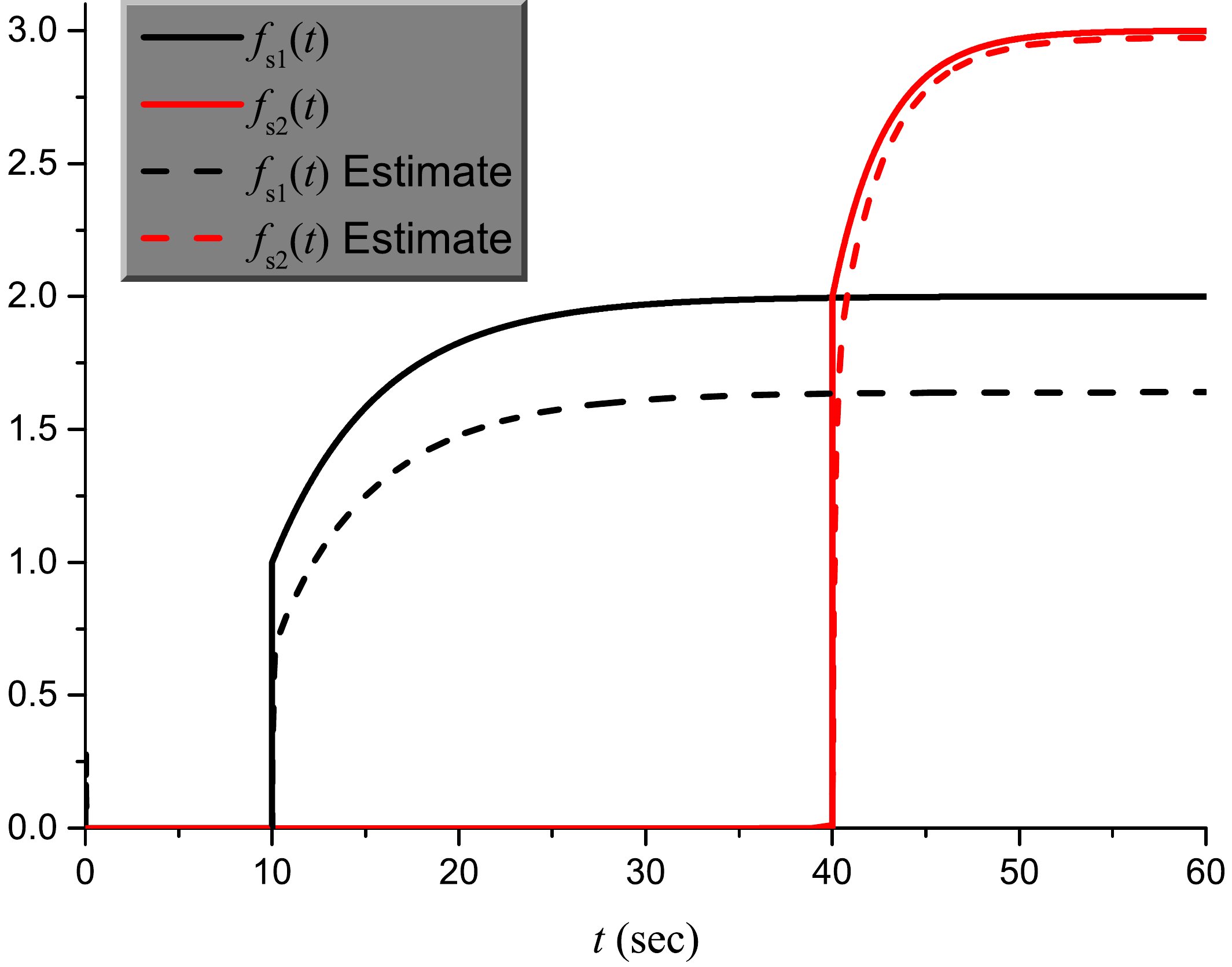}	
				\label{fig:4c}	
			}

		\caption{Estimation results of abnormal source 2. (Incipient source)}
	\end{figure}

   \begin{figure}[!h]
    	\centering
    	\includegraphics[width=7cm]{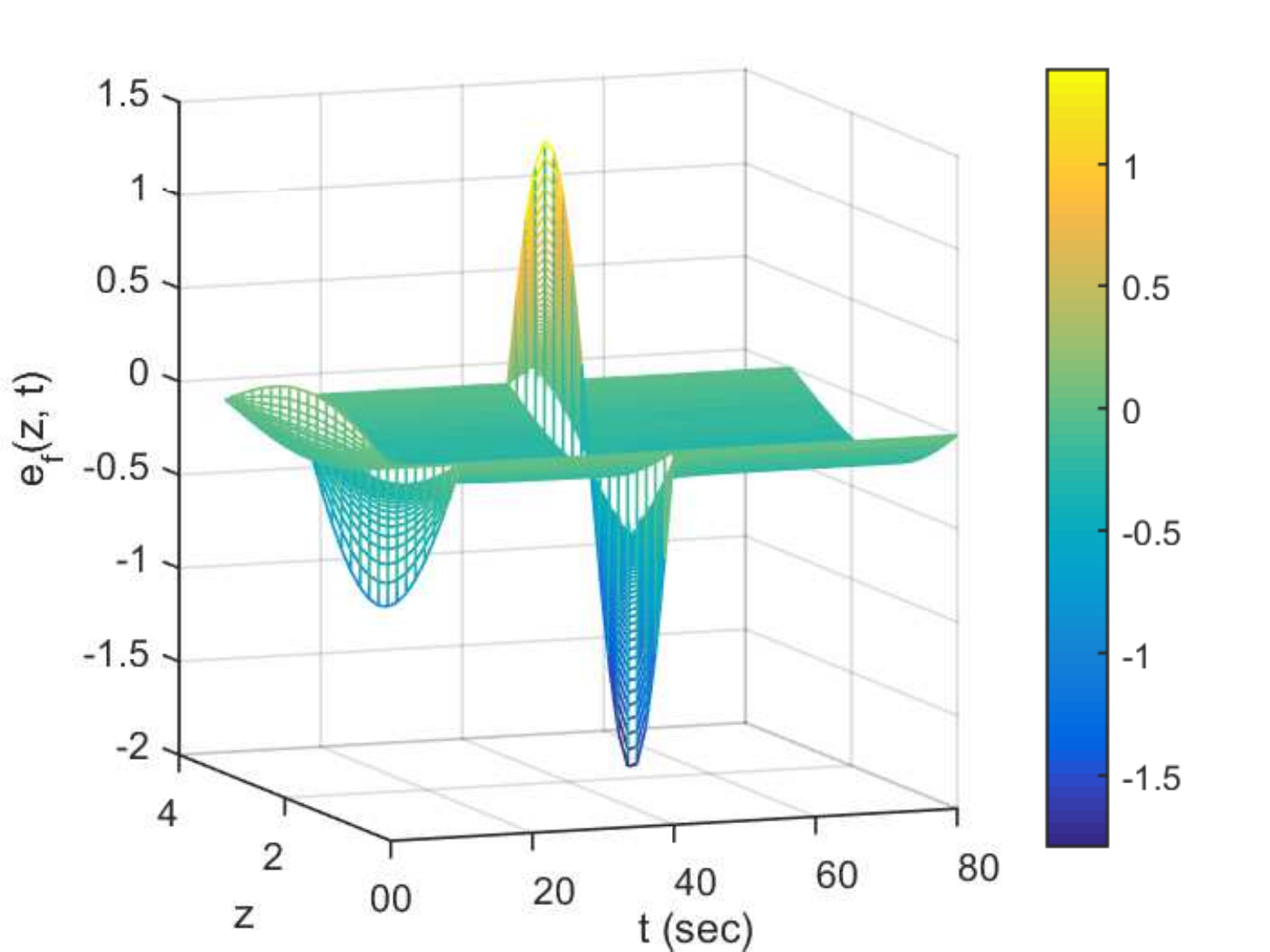}
    	\caption{$e_f(z,t)$ of abnormal source 2. (Incipient source).}
    	\label{fig:ef2}
    \end{figure}

	The simulation results for both abnormal sources are presented in Fig.~\ref{fig:2a}-Fig.~\ref{fig:ef1} and Fig.~\ref{fig:4a}-Fig.~\ref{fig:ef2}, respectively. From Fig.~\ref{fig:2a}, Fig.~\ref{fig:2b}, Fig.~\ref{fig:4a}, and Fig.~\ref{fig:4b}, it can be found that some minor errors exist between the slow subsystem output $\bm{y}_s(t)$ in (\ref{e29}) and the output of the original PDE system $\bm{y}(t)$ in (\ref{e27}), which are caused by the neglecting of the fast subsystem. Meanwhile, the output estimation $\hat{\bm{y}}_s(t)$ in (\ref{e12}) tracks the original PDE system output $\bm{y}(t)$ rapidly under both abnormal sources. In addition, there exist minor errors between $\bm{f}_s(t)$ and its estimation $\hat{\bm{f}}_s(t)$, which are also caused by the truncation error introduced by the neglecting of the fast subsystem. Moreover, the abnormal S-T source estimation $\hat{f}(z,t)$ can track $f(z,t)$ with minor errors, as shown in Fig.~\ref{fig:ef1}, and Fig.~\ref{fig:ef2}, the abnormal source occurring time ($t=10$ (sec) and $40$ (sec)) can be detected for both kinds of sources. To better illustrate the performance of the proposed abnormal S-T source identification approach, the performance index ``RMSE (root of mean squared error)'' is defined as:
	\begin{displaymath}
	\text{RMSE}={(\int {\sum {{e_f}{{(z,t)}^2}dz/\int {dz\sum {\Delta t} } } } )^{1/2}}.
	\end{displaymath}        
	The calculated value of RMSE for abrupt and incipient abnormal source estimation in Fig.~\ref{fig:ef1} and Fig.~\ref{fig:ef2} are $0.2007$ and $0.1919$, respectively. The main reason for the existing of the estimation error $e_f(z,t)$ is the neglecting of the fast subsystem, as discussed in Remark 2. 
		
	For simple illustration, the above S-T source is generated using the eigenfunctions of the spatial operator $\mathcal{A}$ as the basis functions. To further study the effectiveness of the proposed method, consider a general S-T source $f(z,t)$ as follows:
	\begin{displaymath}
	f(z,t)=f(t)b_f(z),
	\end{displaymath}
	where
	\begin{equation*}
	\begin{aligned}
	b_f(z)&= H(z)-H(z-\pi/4),\\
	f(t)&=\left\{\begin{array}{l}{0,\;0 \leqslant t<10(\sec )} \\ {2,\; 10 \leqslant t \leqslant 80(\mathrm{sec})}\end{array}\right.
	\end{aligned}
	\end{equation*}
	$H(\cdot)$ denotes the standard Heaviside function.
	
	To study the effectiveness of the proposed method, $n_y$ point-wise measurements are \textbf{uniformly distributed} in the spatial domain $[0,\pi]$. Moreover, the first $m$ eigenvalues were selected as the \textbf{dominant modes}. The abnormal source estimation results are provided below using the performance index RMSE.
	
	\begin{table}[!h]
		\renewcommand{\arraystretch}{1}
		\caption{Identification results of a general S-T source}
		\centering
        \label{tab:S-T results}
		\setlength{\tabcolsep}{5.5mm}{
			\begin{tabular}{l l l l}
				\hline\hline \\[-2mm]
				\multicolumn{1}{c}{$(m, n_y)$} & \multicolumn{1}{c}{$\bm{\Gamma}$} & \multicolumn{1}{c}{RMSE} & \multicolumn{1}{c}{Ideal RMSE} \\[1.6ex] \hline
  			     (2, 2) & $100*I_2$ & 0.7709 & 0.7497\\ 
                 (2, 3) & $100*I_2$ & 0.7517 & 0.7497  \\ 
                 (3, 3) & $100*I_3$ & 0.6377 & 0.5901\\ 
                 (2, 4) & $100*I_2$ & 0.7518 & 0.7497\\ 
                 (3, 4) & $100*I_3$ & 0.6454 & 0.5901 \\ 
                 (4, 4) & $100*I_4$ & 0.5102 & 0.4286\\ 
				\hline\hline
			\end{tabular}
		}
	\end{table}  

   As shown in Table~\ref{tab:S-T results}, the performance of the proposed method on this S-T source is not as well as that on S-T source generated by the same basis functions. The reason is evident since for this kind of source, the truncation error is relatively large and it requires a larger number of basis functions for accurate approximation. It can also be found in this table that RMSE decreases with the increasing of the number of the dominant modes $m$, which is reasonable. To better illustrate the performance of the proposed method, another performance index named ``Ideal RMSE" is introduced as:
   \begin{equation*}
   \begin{aligned}
   e_{\text{ideal}f}(z,t)&=\boldsymbol{\phi}_{s}^{T}(z) \boldsymbol{f}_{s}(t)-f(z,t),\\
   \boldsymbol{f}_{s}(t)&=<\boldsymbol{\phi}_{s}^{T}(z), b_f(z)>f(t),\\ 
   \operatorname{Ideal\;RMSE}&=\left(\int \sum e_{\text{ideal}f}(z, t)^{2} d z / \int d z \sum \Delta t\right)^{1 / 2},
   \end{aligned}
   \end{equation*}
   which determines a lower bound of RMSE. Comparing RMSE with Ideal RMSE, it can be found that the proposed method attains acceptable results on this S-T source. For those S-T sources generated by the same basis functions, the Ideal RMSE=0 apparently.    
   
   In fact, for a given number of sensors, how to place them as well as choosing the learning rate $\bm{\Gamma}$ to obtain better estimation performance is one interesting multi-objective optimization problem. In this manuscript, the focus is on providing a general framework for solving such an inverse source estimation problem rather than achieving the best estimation performance, which is very challenging itself. We will discuss such problems in subsequent researches.

	\begin{remark}
	   	As the requirements of the modulating functions-based method are too restrictive, it is not fair nor meaningful to compare these two methods' performance on the same problem. To be more specific, on one hand, if these requirements of the modulating functions-based method cannot be met, it will not work definitely; On the other hand, since we aim to release these restrictions for industrial applications in this paper, the significance of the proposed method would be largely decreased if these requirements can be met,   regardless of the comparison results.	
	\end{remark}
   	
	\section{Conclusion}\label{sec:Conclusion}
	In this paper, the abnormal S-T source identification for a class of linear parabolic DPSs is first investigated. An inverse S-T model for abnormal source identification is developed, which consists of an adaptive state observer for source identification and an adaptive source estimation algorithm. Theoretic analysis is provided to guarantee the convergence of the abnormal S-T source estimation error. Finally, numerical simulations on a heat rod with an abnormal S-T source are presented to evaluate the performance of the proposed method. Future researches will be expanded to the abnormal S-T source identification for non-linear DPSs.


	\appendix
	\section{Model reduction for parabolic DPSs} \label{App:A}    
	First, define the operator $\mathcal{A}$ in ${\mathcal{H}}$ as: 
	\begin{equation*}
	\begin{gathered}
	\mathcal{A} {x} = a_1\frac{{\partial {x}}}
	{{\partial z}} + a_2\frac{{{\partial ^2} {x}}}
	{{\partial {z^2}}}+a_3x, {x} \in \mathcal{S}(\mathcal{A})\mathop  = \limits^\Delta  \{  {x} \in \mathcal{H}\\
	,\text{and the conditions in (\ref{e2}) hold}\} .
	\end{gathered}
	\end{equation*}
	For the above spatial operator $\mathcal{A}$, we have the following eigenvalue problem: 
	\begin{equation*}
	\mathcal{A}{\phi _j}(z) = {\lambda _j}{\phi _j}(z),\;j = 1,\cdots,\infty, 
	\end{equation*}   
	where $\phi_j$ denotes the eigenfunction and $\lambda_j$ denotes the corresponding eigenvalue. Define $\sigma (\mathcal{A})$ as the set of all eigenvalues of $\mathcal{A}$, that is $\sigma (\mathcal{A}) = \{ {\lambda _1},\cdots,\}.$ The following assumptions state that $\sigma (\mathcal{A})$ can be partitioned into a finite-dimensional part which consists of $m$ slow eigenvalues and a stable infinite-dimensional complement consists of the remaining fast eigenvalues and that the gap between the slow and fast eigenvalues is large.
	
	\begin{assumption}\emph{
		\hfill \\ 
		1. $Re\{\lambda_1\} \geqslant Re\{\lambda_2\} \geqslant\cdots \geqslant Re\{\lambda_j\} \geqslant\cdots,$ where $Re\{\lambda_j\}$ denotes the real part of $\lambda_j$.\\
		2. $\sigma (\mathcal{A})$ can be partitioned as $\sigma (\mathcal{A})=\sigma_1 (\mathcal{A})+\sigma_2 (\mathcal{A})$, where $\sigma_1 (\mathcal{A})$ consists of the first $m$ (with $m$ finite) eigenvalues, that is $\sigma_1 (\mathcal{A})=\{ {\lambda _1},\cdots,{\lambda _m}\}$, and ${\left| {Re \{ {\lambda _1}\} } \right|}/
		{{\left| {Re \{ {\lambda _m}\} } \right|}} = O(1).$\\
		3. $Re\{\lambda_{m+1}\}<0$ and ${\left| {Re \{ {\lambda _m}\} } \right|}/
		{{\left| {Re \{ {\lambda _{m+1}}\} } \right|}} = O(\varepsilon),$ where $\varepsilon<1$ is a small positive number.}
	\end{assumption}

 Based on the theorem of separation of variables~\cite{maccluer1994boundary} and applying Galerkin's method~\cite{christofides2012nonlinear}, the following infinite-dimensional system modeled by ODE is obtained:
	\begin{equation} \label{e8}
	\begin{aligned}
	{{\dot {\bm{x}}}_s}(t) &= {{\bm{A}}_s}{\bm{x}_s}(t)+{\bm{B}_{u,s}}\bm{u}(t)+\bm{f}_s(t), \hfill \\
	{{\dot {\bm{x}}}_f}(t) &= {{\bm{A}}_f}{\bm{x}_f}(t)+{\bm{B}_{u,f}}\bm{u}(t)+\bm{f}_f(t), \hfill \\
	\bm{y}(t) &= {\bm{C} _s}{\bm{x}_s}(t) + {\bm{C}_f}{\bm{x}_f}(t)=\bm{y}_s(t)+\bm{y}_f(t), \hfill \\ 
	\end{aligned}
	\end{equation}
	with the initial conditions:
	\begin{equation*}
	\bm{x}_s(0) =  \mathcal{P}_s\bm{x}_0(z), \bm{x}_f(0) =  \mathcal{P}_f\bm{x}_0(z), 
	\end{equation*} 
	where
	\begin{equation*}
	\begin{gathered}
	{\bm{A}}_s =\mathcal{P}_s\mathcal{A}{\bm{\phi} ^T}(z)= diag\{ {\lambda _1},\cdots,{\lambda _m}\}, \hfill \\
	{\bm{A}}_f =\mathcal{P}_f\mathcal{A}{\bm{\phi} ^T}(z)= diag\{ {\lambda _{m+1}},\cdots,{\lambda _\infty}\}, \hfill \\
	{\bm{B}_{u,s}} =  \mathcal{P}_s k_u\bm{b}_u^T(z),\;{\bm{B}_{u,f}} =  \mathcal{P}_f k_u\bm{b}_u^T(z), \hfill \\
	{\bm{f}_s(t)} =  \mathcal{P}_s f(z,t),\;{\bm{f}_f(t)} =  \mathcal{P}_f f(z,t), \hfill \\
	{\bm{C}_s} = \int_{{\alpha _1}}^{{\alpha _2}} {\bm{c}(z){k_y}\bm{\phi} _s^T(z)dz} ,{\bm{C}_f} = \int_{{\alpha _1}}^{{\alpha _2}} {\bm{c}(z){k_y}\bm{\phi} _f^T(z)dz}. \hfill \\
	\end{gathered}
	\end{equation*}
	
	The derivations of this infinite-dimensional ODE systems are neglected since they are trivial, interesting readers may seek for more details in~\cite{christofides2012nonlinear,wu2011robust}.  
	
	Using $\varepsilon  = \frac{{\left| {\operatorname{Re} \{ {\lambda _1}\} } \right|}}
	{{\left| {\operatorname{Re} \{ {\lambda _{m + 1}}\} } \right|}}$ and multiplying the $\bm{x}_f$-subsystem by $\varepsilon$ yields the singular perturbation model:
	\begin{equation} \label{e9}
	\begin{aligned}
	{{\dot {\bm{x}}}_s}(t) &= {{\bm{A}}_s}{\bm{x}_s}(t)+{\bm{B}_{u,s}}\bm{u}(t)+\bm{f}_s(t), \hfill \\
	\varepsilon{{\dot {\bm{x}}}_f}(t) &= {{\bm{A}}_{f\varepsilon}}{\bm{x}_f}(t)+\varepsilon{\bm{B}_{u,f}}\bm{u}(t)+\varepsilon \bm{f}_f(t), \hfill \\
	\end{aligned}
	\end{equation}   
	where ${\bm{A}}_{f\varepsilon}=\varepsilon{\bm{A}}_f$. 
	
	Then the singular perturbation theory~\cite{khalil2002nonlinear} can be applied. Using the fast time-scale $\tau  = t/\varepsilon$ and set $\varepsilon=0$, we can derive the following infinite-dimensional fast subsystem directly from model (\ref{e9}):
	\begin{equation} \label{e10}
	\frac{{d{\bm{x}_f}(\tau )}}
	{{d\tau }} = {{\bm{A}}_{f\varepsilon }}{\bm{x}_f}(\tau ).
	\end{equation}
	Since $Re\{\lambda_{m+1}\}<0$ and referring to the definition of $\varepsilon$, it can be concluded that (\ref{e10}) is globally exponentially stable. Maintaining $\varepsilon=0$ in (\ref{e9}) and considering the non-singularity of ${\bm{A}}_{f\varepsilon}$, it can be obtained that $\bm{x}_f(t)=0$. Substituting $\bm{x}_f(t)=0$ into (\ref{e9}), it can be obtained that:
	\begin{equation*} 
	\begin{aligned}
	{{\dot {\bm{x}}}_s}(t) &= {{\bm{A}}_s}{\bm{x}_s}(t)+ {\bm{B}_{u,s}}\bm{u}(t)+\bm{f}_s(t), \hfill \\
	{\bm{y}}_s(t) &= {\bm{C}_{s}{\bm{x}_s}(t)}. \hfill \\
	\end{aligned}
	\end{equation*}

%
%
%
	\section*{Acknowledgment}

	The authors would like to thank the Editor-in-Chief, Associate Editor and anonymous reviewers for their valuable comments based on which the presentation of this paper has been improved.

	\ifCLASSOPTIONcaptionsoff
	\newpage
	\fi

	
	
	%
	%
	%
	
	\bibliographystyle{bib/IEEEtran}
	\bibliography{bib/IEEEabrv,bib/mybib}
	%
%
\vspace{0cm}
\begin{IEEEbiography}[{\includegraphics[width=1in,height=1.25in,clip,keepaspectratio]{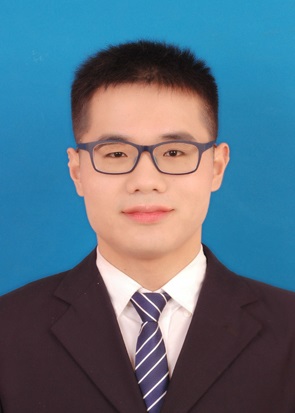}}]
	{Yun Feng} received his B.E. degree in automation and M.S. degree in control theory and control engineering both from the Department of Automation, Wuhan University, Wuhan, China, in 2014 and 2017, respectively. He is currently pursuing the Ph.D. degree with the Department of Systems Engineering and Engineering Management, City University of Hong Kong, Hong Kong. From July to November 2019, he was a visiting student at the Institute for Automatic Control and Complex Systems (AKS), University of Duisburg-Essen, Germany.  
	
	His current research interests include fault diagnosis of distributed parameter systems and computational intelligence.\\

\end{IEEEbiography}

\vspace{1cm}
\begin{IEEEbiography}[{\includegraphics[width=1in,height=1.25in,clip,keepaspectratio]{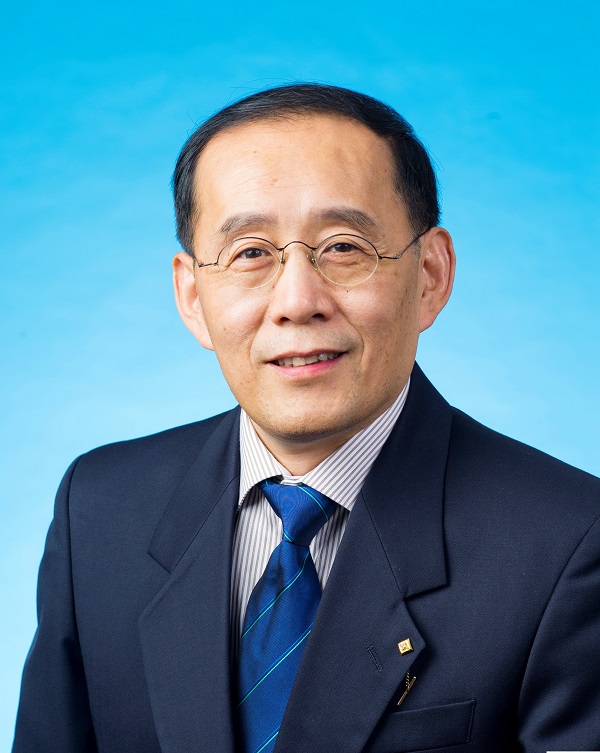}}]
	{Han-Xiong Li} (S'94-M'97-SM'00-F'11) received his B.E. degree in aerospace engineering from the National University of Defense Technology, China in 1982, M.E. degree in electrical engineering from Delft University of Technology, The Netherlands in 1991, and Ph.D. degree in electrical engineering from the University of Auckland, New Zealand in 1997. 
	
	He is a professor in the Department of SEEM, City University of Hong Kong. He has a broad experience in both academia and industry. He has authored 2 books and about 20 patents, and published more than 200 SCI journal papers with h-index 45 (web of science). His current research interests include process modeling and control, system intelligence, distributed parameter systems, and battery management system. 
	
	Dr. Li serves as Associate Editor for IEEE Transactions on SMC: System, and was associate editor for IEEE Transactions on Cybernetics (2002-2016) and IEEE Transactions on Industrial Electronics (2009-2015). He was awarded the Distinguished Young Scholar (overseas) by the China National Science Foundation in 2004, a Chang Jiang professorship by the Ministry of Education, China in 2006, and a national professorship in China Thousand Talents Program in 2010. He serves as a distinguished expert for Hunan Government and China Federation of Returned Overseas Chinese. He is a Fellow of the IEEE.\\ 
	
\end{IEEEbiography}
	
	
	

\end{document}